\documentclass[fleqn]{cas-dc}  
\pdfoutput=1

\usepackage[authoryear]{natbib}

\usepackage{lineno}
\usepackage{amsmath,amssymb}

\usepackage{journals} 
\usepackage{booktabs}
\usepackage{longtable}
\usepackage{lineno}
\usepackage{ulem}

\newcommand{\tmelt}{T_\mathcal{M}}

\newcommand{\phase}{\phi}
\newcommand{\phasepara}{a}

\newcommand{\volL}{\mathcal{V}_L}
\newcommand{\volS}{\mathcal{V}_S}
\newcommand{\EL}{E_\text{eff}}
\newcommand{\RaL}{Ra_\text{eff}}
\newcommand{\etaL}{\eta_\text{eff}}
\newcommand{\hL}{h_L}
\newcommand{\NuL}{Nu_\text{eff}}
\newcommand{\hS}{h_S}
\newcommand{\rmeltmean}{\xi_\mathcal{M}}
\newcommand{\rmelt}{r_\mathcal{M}}
\newcommand{\tautopo}{\tau_\xi}
\newcommand{\qhl}{\mathcal{Q}^{h/l}}
\newcommand{\hshl}{h_S^{h/l}}

\DeclareMathOperator{\argsinh}{argsinh}
\DeclareMathOperator{\argmax}{\arg\!\max}

\def\vec#1{\ensuremath{\mathchoice{\mbox{\boldmath$\displaystyle#1$}}
{\mbox{\boldmath$\textstyle#1$}}
{\mbox{\boldmath$\scriptstyle#1$}}
{\mbox{\boldmath$\scriptscriptstyle#1$}}}}

\begin{document}
\let\WriteBookmarks\relax
\def\floatpagepagefraction{1}
\def\textpagefraction{.001}



\shorttitle{Rotating convection with a melting boundary}

\title[mode=title]{
Rotating convection with a melting boundary: an application to the icy moons}

\shortauthors{T.~Gastine and B.~Favier}
\author[1]{T.~Gastine}[orcid=0000-0003-4438-7203]
\author[2]{B.~Favier}[orcid=0000-0002-1184-2989]

\affiliation[1]{organization={Universit\'e Paris Cit\'e, Institut de physique du globe de Paris, UMR 7154 CNRS},addressline={1 rue Jussieu},city={F-75005, Paris},country=France}
\affiliation[2]{organization={Aix Marseille Univ, CNRS, Centrale Med, IRPHE},city={Marseille},country=France}

\cormark[1]
\cortext[1]{Corresponding author. E-mail: 
\href{mailto:gastine@ipgp.fr}{gastine@ipgp.fr}}


\begin{abstract}
A better understanding of the ice-ocean couplings is required to better characterise the hydrosphere of the icy moons.
Using global numerical simulations in spherical geometry, we have investigated here the interplay between rotating convection and a melting boundary.
To do so, we have implemented and validated a phase field formulation in the open-source code \texttt{MagIC}. 
We have conducted a parameter study varying the influence of rotation, the vigour of the convective forcing and the melting temperature.
We have evidenced different regimes akin to those already found in previous monophasic models in which the mean axisymmetric ice crust transits from pole-ward thinning to equator-ward thinning with the increase of the rotational constraint on the flow.
The derivation of a perturbative model of heat conduction in the ice layer enabled us to relate those mean topographic changes to the underlying latitudinal heat flux variations at the top of the ocean.
The phase change has also been found to yield the formation of sizeable non-axisymmetric topography at the solid-liquid interface with a typical size close to that of the convective columns.
We have shown that the typical evolution timescale of the interface increases linearly with the crest-to-trough amplitude and quadratically with the mean melt radius.
In the case of the largest topographic changes, the convective flows become quasi locked in the topography due to the constructive coupling between convection and ice melting.
The tentative extrapolation to the planetary regimes yields  $\mathcal{O}(10^2-10^3)$ meters for the amplitude of non-axisymmetric topography at the base of the ice layer of Enceladus and $\mathcal{O}(10^3-10^4)$ meters for Titan.
\end{abstract}


\maketitle

\section{Introduction}

The presence of liquid water oceans has been evidenced on at least four
icy satellites of our solar system: Europa, Ganymede, Titan and Enceladus \citep[e.g.][]{Soderlund20}.
The determination of the structural properties of their hydrosphere is one of the main
objective of the ongoing ESA's JUICE and NASA's Europa Clipper missions \citep{VanHoolst24,Roberts23}.
In this context, it is necessary to better understand the dynamics of the subsurface oceans and their interplay with their overlying ice shelves. Oceans of the icy moons are subjected to
buoyancy forcing of thermal and solutal origins as well as mechanical forcing such as libration 
or tides \citep{Soderlund23}. 
The aim of this study is to focus on thermally-driven flows and their interaction with the
overlying ice crust.

Using global numerical simulations, previous parameter studies have delineated several dynamical physical regimes of rotating 
convection in spherical geometry \citep{Gastine16,Long20}. The relative influence
of rotation is usually categorised in terms of boundary regimes defined
as combinations of the dimensionless governing parameters relevant to the fluid problem, namely the
Ekman number $E$, the Rayleigh number $Ra$ and the Prandtl number $Pr$ (to be defined below).
In the rapidly-rotating regime, the convective flow is governed by a quasi-geostrophic
balance which imposes a strong alignment of the convective flows with
the rotation axis.
Multiple alternating zonal jets can form \citep{Ashkenazy21,Bire22}, reaching
a typical width close to the so-called Rhines scale \citep{Rhines75,Cabanes24}. On the other hand, when the 
rotational constraint drops, the flow becomes three-dimensional and the mean zonal flows occupies 
a decreasing fraction of the kinetic energy content \citep{Gastine13,Yadav16}. 

In addition to these zonal flow changes, the heat flux pattern also evolves with the rotational 
constraint \citep{Gastine23}: heat flux at the outer boundary of the ocean may either peak near 
the equator (the so-called ``equatorial cooling'' regime) or close to the poles (the ``polar cooling'' regime) \citep{Soderlund19,Amit20}.
The exact parameter combination of $Ra$, $E$ and $Pr$ which governs the transition between these 
two regimes is still a matter of debates and depends on the nature of the mechanical boundary 
condition \citep{Kvorka22}, the radius ratio of the subsurface ocean \citep{Bire22} and
the coupling with the mean zonal flows \citep{Gastine23}.
Using stress-free boundary conditions, \citet{Kvorka22} report a transition between
equatorial and polar cooling when $Ra\,E^{12/7}Pr^{-1} \approx 1$, while 
\citet{Hartmann24} rather favour $Ra\,E^{3/2} \sim 1$ when rigid boundaries are employed 
\citep[see also][]{Bire22}.
Although the uncertainties remain sizeable, current estimates of the dimensionless numbers
of the sub-glacial oceans of the icy moons would most likely place them in 
a weakly-rotating convection regime  rather prone to polar cooling \citep{Soderlund19,Lemasquerier23}.

In the case of a conducting ice layer, 
an increase of the heat flux at the solid-liquid interface goes along  with a thinning of the overlying ice, such that the ice shell thickness is expected to be anti-correlated with the heat flux pattern atop the ocean \citep{Kvorka18,Kihoulou23}.
This hypothesis prompts several authors to interpret observations such as the chaos terrain 
in Europa's equatorial regions \citep{Soderlund14}, the polar depressions on Titan \citep{Kvorka18}
or the poleward thinning of the ice crust on Enceladus \citep{Cadek19} in terms of the relative changes of the underlying convective heat flux \citep[see also][]{Kvorka24}.

Recent large eddy simulations which account for the local changes of the melting temperature
along the solid-liquid interface however challenged this interpretation 
\citep{Ashkenazy21,Kang22}. Because of the melting point dependence on pressure
\citep[e.g.][]{Labrosse18}, the $20$~km ice thickness variation
between the South pole and the equator of Enceladus \citep{Hemingway19} 
would approximately yield a $0.2$~K 
latitudinal difference at the base of the ice crust \citep{Lawrence24}. 
This thermal gradient could drive large-scale baroclinic flows and 
hamper the convective transport in the polar regions 
\citep{Kang23}, hence questioning the relation between the heating pattern and the ice geometry.
Using 2-D Cartesian models in which the oceanic flows are driven by the latitudinal variations of the melting temperature at the solid-liquid interface, \citet{KangJansen22} suggest that the heat transport is more efficient
on larger icy moons because of their higher gravity. Corresponding equilibrated ice shells are then expected to become flatter for increasing body sizes.
Variations in salinity due to melting or freezing of water \citep{Ashkenazy21,Wong22} 
or heat flux heterogeneities at the ocean's base or at the moon's surface 
\citep{Terra23,Lemasquerier23} are additional physical ingredients likely
to modify the interplay between oceanic flows and the overlying ice layer.
These competing interpretations also stem from the challenge to draw reliable scaling laws
able to bridge the gap between the control parameters accessible to current
numerical models and the relevant planetary regime \citep{Jansen23,Cabanes24}.

In current numerical simulations dedicated to the influence of ice thickness variations
on the oceanic flows, the ice crust is however assumed to be static
\citep[e.g.][]{Kang22,Kang23}. This prohibits the dynamical generation of topographic
features associated with the turbulent flows. The purpose of the present study 
is precisely to focus on the interplay between rotating convection in spherical geometry
and generation of topography by melting or freezing of the overlying ice crust.
To do so, we consider a phase field formulation which allows
to model a two-phase fluid problems on a single fixed-grid domain \citep[e.g.][]{Beckermann99}.
This type of approach is a smoothed approximation of the phase 
change which greatly simplifies the numerical implementation and has been shown to converge towards the exact moving-boundary formulation \citep[e.g.][]{Hester20}. 
Among many examples, let us cite the 
successful application of this method to the study of 
Rayleigh-B\'enard convection (RBC) with a melting boundary 
in Cartesian geometry \citep{Favier19,Yang23}, the generation of topography in a turbulent
shear flow \citep{Couston21,Perissutti24}, the formation of pinnacles or scallops in melting ice \citep{Weady22}, or the influence of the aspect ratio of an iceberg on its melting \citep{Hester21}.
All of these examples showed complex dynamical interactions between the convective flow pattern and the morphology of the phase change interface which can only be captured with an explicit treatment of its dynamics.
To date, there is a scarce number of studies dedicated to rotating convection with a dynamical phase change boundary. 
Using numerical simulations in Cartesian geometry, \citet{Ravichandran21} examine the 
melting of a solid layer above a convecting liquid domain rotating about its vertical
axis. They report on quasi-steady states in which the convective columns are locked
in the topographic troughs and crests of the solid-liquid interface.

To our best knowledge, our present study is the first one to consider 
a phase field formulation applied to convection in a global spherical geometry. 
We would like to stress that this work should be regarded as a first
incremental step  in the process of improving our understanding of the
dynamical interplay between oceanic flows and the overlying ice layers. 
As such, several effects relevant to the sub-glacial oceans of the icy
moons, such as salinity, creep of ice or the pressure dependence
of melting temperature, have been ignored in the present study.

The paper is organised as follows. The physical model including the phase field
formulation and its numerical implementation and validation is discussed in
\S~\ref{sec:model}. Results of the numerical simulations are described
in \S~\ref{sec:results}, where we have split the analysis in terms of a mean-field
approach, discussing first the axisymmetric topography and then
the non-axisymmetric roughness. Tentative geophysical estimates
are discussed in \S~\ref{sec:geophy} before concluding in \S~\ref{sec:conclu}.

\section{Model and methods}
\label{sec:model}
\subsection{A phase field model}

We consider a spherical shell gap of inner radius $r_i$ and outer 
radius $r_o$ with $d=r_o-r_i$ and $\eta=r_i/r_o$ which rotates with a constant rotation rate $\Omega$ about the $z$-axis. Convection is enforced by maintaining a fixed temperature contrast $\Delta T$ between both boundaries.
We explore the uneven generation of topography associated with the freezing and melting which occurs at the fluid-solid interface for a melting temperature comprised between the imposed temperatures at the top and bottom boundaries.
The solid phase is motionless and located in the outer part of the spherical volume.
In the following, we consider a dimensionless formulation of the Navier-Stokes equations under the Boussinesq approximation with a constant kinematic viscosity $\nu$ and thermal diffusivity $\kappa$.
We employ the viscous diffusion time $d^2/\nu$ as the reference time scale and the imposed temperature contrast $\Delta T$ as the 
temperature scale.
To model the phase changes, we adopt the phase field formulation by \citet{Beckermann99} combined with a volume-penalization technique \citep{Angot99,Hester21a}.
Practically, this method involves the time integration of a continuous scalar quantity $\phase$ which continuously varies from $0$ in the liquid phase to $1$ in the solid phase.
A small dimensionless parameter $\epsilon=\lambda/d$, usually termed 
the Cahn number, then defines the ratio between the microscopic thickness of the transition between the two phases $\lambda$ and the macroscopic domain size which is here the shell gap $d$.
Phase field methods represent a smoothed formulation of phase changes which are easier to implement numerically, especially when using pseudo-spectral methods, and converge to the exact moving boundary formulation in the limit of vanishing $\epsilon$ 
\citep[e.g.][]{Caginalp86}.
Using the 
model recently derived by \cite{Hester20}, and used in many subsequent studies \citep[e.g.][]{Weady22,Yang23}, the governing equations for the velocity field $\vec{u}$, the temperature $T$ 
and the phase field $\phase$ then read
\begin{equation}
 \vec{\nabla}\cdot\vec{u} = 0,
 \label{eq:divu}
\end{equation}
\begin{equation}
 \dfrac{\partial \vec{u}}{\partial t} +\vec{u}\cdot\vec{\nabla}\vec{u}
 +\dfrac{2}{E}\vec{e}_z\times \vec{u}= 
-\vec{\nabla} p + \dfrac{Ra}{Pr} g \vec{e}_r + \vec{\nabla^2}\vec{u}
-\dfrac{1}{\tau_p \epsilon^2}\phase \vec{u}\,,
\label{eq:NS}
\end{equation}
\begin{equation}
 \dfrac{\partial T}{\partial t}+\vec{u}\cdot\vec{\nabla}T = \dfrac{1}{Pr} 
\nabla^2 T+St\dfrac{\partial \phase}{\partial t}\,,
\label{eq:temp}
\end{equation}
\begin{equation}
 \dfrac{5}{6}St Pr\dfrac{\partial 
\phase}{\partial 
t} = \phasepara \nabla^2 \phase 
-\dfrac{1}{\epsilon^2}\,\phase(1-\phase)\left[\phasepara(1-2\phase)+T-\tmelt
\right]\,.
 \label{eq:phase}
\end{equation}
In the above expressions $\vec{e}_r$ and $\vec{e}_z$ denote the unit vectors in the radial and axial directions while $g=r/r_o$ is the dimensionless gravity profile.
$\tmelt$ is the dimensionless melting temperature, with $0<\tmelt<1$ delimiting the regimes for which both liquid and solid phases are present within the spherical shell.
In the limit of vanishing phase-change interface $\epsilon$, Eq.~\eqref{eq:phase} effectively imposes 
the so-called Stefan conditions \citep{Huppert90,Worster00}, which relate the interface velocity to the heat flux differences at the solid-liquid interface and ensure that the phase change interface is isothermal according to
\begin{align}
\label{eq:stefan}
 \vec{n} \cdot \left [ \vec{\nabla} T^{(S)}-\vec{\nabla}T^{(L)}\right] & = St Pr 
\,\vec{v}\cdot\vec{n}, \\
\label{eq:stefant}
T=\tmelt,
\end{align}
where $\vec{n}$ is the unit vector normal to the interface, $\vec{v}$ is the interface velocity and the superscripts $^{(S)}$ and $^{(L)}$ respectively correspond to the solid and liquid phases.
Within the phase field formulation, the interface is implicitly defined by the isosurface $\phase=1/2$.

For simplicity, our model assumes that thermal diffusivity and density are the same in both phases and we neglect any compositional effects (such as the dependence of the melting temperature on salinity for example).
Density could increase by about $20\%$ in Ganymede's hydrosphere, most of the contrast being accommodated in the fluid phase with values ranging from
$950~$kg/m$^3$ in the ice crust to about $1200~$kg/m$^3$ at the ocean's base \citep[e.g.][their Table~2]{Journaux20}. Bearing in mind the possible unknowns in the transport properties in the interiors of the icy moons, thermal diffusivity of ice is expected to be larger than the one of liquid water by about a factor $5-10$ covering the range $\kappa \sim 10^{-7}-10^{-6}$~m$^2$/s \citep[e.g.][]{Abramson01,Vance18}.
Additionally, the melting temperature does not depend on pressure, an effect that is potentially important for many deep fluid systems from sub-glacial lakes on Earth \citep[e.g.][]{Couston21} to oceans on icy satellites \citep[e.g.][]{Labrosse18,Kang23}.
While such thermobaric and salinity effects can be incorporated in our approach \citep[see][for recent examples involving salinity]{Hester20,Yang23a}, we choose to ignore them in this preliminary study and defer the question of their contributions to future studies.

The set of equations (\ref{eq:divu}-\ref{eq:phase}) is governed by four physical dimensionless numbers: the Ekman number $E$, the Rayleigh number $Ra$, the Prandtl number $Pr$ and the Stefan number $St$ expressed by
\begin{equation}
\label{eq:params}
 E=\dfrac{\nu}{\Omega d^2},\ Pr=\dfrac{\nu}{\kappa},\ 
 Ra=\dfrac{\alpha g_o \Delta T d^3}{\nu\kappa},\ 
 St=\dfrac{\mathcal{L}}{c_p \Delta T}\,.
\end{equation}
In the above expressions, $g_o$ denotes the gravity at the outer boundary, $c_p$ corresponds to the heat capacity, $\alpha$ to the thermal expansion coefficient and $\mathcal{L}$ to the latent heat per unit mass associated with the solid-liquid transition. 

The phase field  formulation and the penalty method involve several additional dimensionless parameters compared to more classical monophasic models.
The phase-change interface thickness controlled by the Cahn number $\epsilon$ is chosen to be smaller than the thickness of the Ekman boundary layers, which is the smallest relevant physical scale close to the phase change interface in our problem.
This choice, while very constraining numerically, was motivated by systematic convergence tests carried out by gradually reducing $\epsilon$.
While larger values of the interface thickness might be relevant for modelling mushy layers \citep{LeBars06}, we nevertheless opt for this conservative approach in order to ensure a meaningful comparison with monophasic simulations.
In addition, our model also involves two other dimensionless quantities: a volume-penalization parameter $\tau_p$ 
and a parameter related to the phase field model, $\phasepara=\gamma/\lambda \Delta T$, where $\gamma$ 
expresses the curvature dependence of the melting temperature
expressed in units of $K m$ \citep[see][]{Beckermann99}.
This curvature effect is however very small in practice and only relevant when considering dendritic 
growth or other microscopic phenomena.
To satisfy the standard isothermal Stefan
condition~\eqref{eq:stefant}, $\gamma$ must remain as small as possible, which is the case provided that the
dimensionless parameter $a$ remains of order one while $\epsilon \ll 1$ \citep{Hester20}.
In practice, following \citet{Couston21} and \citet{Yang23}, we adopt $a=1$ for all the simulations considered in 
this study.
Concerning the penalisation term $-\phi\vec{u}/(\tau_p\epsilon^2)$ in Eq.~\eqref{eq:NS}, its role is to 
exponentially attenuate the velocity inside the solid phase, effectively treating it as a porous medium using a 
Darcy-Brinkman model \citep{LeBars06}.
Although an optimal value for the parameter $\tau_p$ can be derived in simple configurations \citep{Hester21}, we follow a different approach in our setup which involves rotation and buoyancy forces.
The value of $\tau_p$ is practically adjusted for each simulation to ensure that the kinetic energy density of the solid phase remains smaller than that of the fluid phase by a factor of at least $10^{-4}$.

Considering a simplified model without rotation and buoyancy,
\cite{Hester20} report on the second-order convergence of the phase-field equation~\eqref{eq:phase}
with respect to the interface thickness $\epsilon$ towards the actual Stefan problem~\eqref{eq:stefan}-\eqref{eq:stefant}.
In Appendix~\ref{sec:bench}, we examine the convergence properties of the phase field formulation by defining a weakly-supercritical quasi-stationary benchmark for rotating convection in spherical shell with a melting boundary.
The convergence of our phase field formulation is found to be more complex than anticipated by \citet{Hester20}, with a gradual change in slope between first and second orders, similar to previous findings by \citet{Favier19}.
This difference is attributed to the influence of rotation and buoyancy forces, two physical effects overlooked by \citet{Hester20} in their analysis.

We adopt the spherical coordinates $(r,\vartheta,\varphi)$ and assume rigid 
mechanical boundary conditions held at constant temperatures with 
$\vec{u}(r=r_i)=\vec{u}(r=r_o)=0$ and $T(r=r_i)=1$ and $T(r=r_o)=0$.
We assume Dirichlet boundary conditions on the phase field with the outer 
(inner) boundary in the solid (liquid) phase, i.e. $\phase(r=r_o)=1$ and 
$\phase(r=r_i)=0$.

\subsection{Numerical methods}

The numerical models considered in this study have been computed using the open source code \texttt{MagIC}\footnote{\url{https://github.com/magic-sph/magic}} combined with the \texttt{SHTns}\footnote{\url{https://gricad-gitlab.univ-grenoble-alpes.fr/
schaeffn/shtns}} library to handle the spherical harmonic transforms \citep[see][]{Schaeffer13}. 
To enforce the divergence-free constraint on the velocity 
field, $\vec{u}$ is expressed in terms of poloidal and toroidal potentials following
\[
 \vec{u}= \vec{u}_P+\vec{u}_T=\vec{\nabla}\times(\vec{\nabla}\times W 
\vec{e}_r) + 
\vec{\nabla}\times Z \vec{e}_r\,.
\]
The quantities $W$, $Z$, $p$, $T$ and $\phase$ are expanded in spherical harmonics up to a degree and order $\ell_\text{max}$ in the horizontal directions and in Chebyshev polynomials up to a degree $N_r$ in the radial direction.
We employ the pseudo-spectral formulation of the equations 
introduced by \citet{Glatzmaier84} in which the nonlinear terms are handled in the physical grid space and then transformed
to the spectral space using spherical harmonic transforms \citep[for details, see for instance][]{Christensen15}.

The equations are advanced in time using implicit-explicit (IMEX) time schemes which handle the non-linear terms and the Coriolis force explicitly and the linear terms implicitly.
The explicit treatment of the additional volume penalization term entering Eq.~\eqref{eq:NS} yields an extra constraint on 
the time step size $\delta t$ \citep{Kolo09} compared to classical convection problems, such that
\begin{equation}
 \max (\delta t) < C\,\tau_p \epsilon^2,
 \label{eq:courant_phase}
\end{equation}
where $C$ is a Courant factor that depends on the IMEX time integrator. 
In this study, we employ the third order backward difference 
scheme (SBDF3) for most of the numerical simulations and the 
third-order diagonally-implicit Runge-Kutta schemes ARS343 from 
\citet{Ascher97} for the most turbulent configurations \citep[for a comparison for convection problems, see][]{Gopinath22}.

The phase field formulation involves large gradients of $\phase$ localised around the melting radius $\rmelt$. 
Here we employ a Chebyshev collocation technique combined with mapping functions capable of handling such steep fronts. Practically, the radial grid points are defined by
\[
 r_k=\dfrac{1}{2}(r_o-r_i)\mathcal{F}(x_k)+\dfrac{1}{2}(r_o+r_i) \quad \textrm{with} \quad k\in[1,N_r]\,
\]
where $x_k=\cos[(k-1)\pi/(N_r-1)]$ are the Gauss-Lobatto (GL) nodes and $\mathcal{F}$ a mapping function. Of particular interest are the mapping functions which allow to refine the resolution around localised regions of rapid variations, such as the ones introduced by \citet{Bayliss92} and \citet{Jafari15} (hereafter respectively BT and JVH).
The latter one has in addition been designed to minimize Gibbs phenomena when solving Allen-Cahn type equations.
For the definitions of the mapping functions, see the Appendix~\ref{sec:mappings}.
The melt radius $\rmelt$ is however susceptible to substantially vary over the horizontal directions in the case of large topography, making the tuning of the mapping parameters difficult. In that case, it is also worthy considering the mapping by \citet{Kosloff93} (hereafter KTE) which simply redistributes the native Gauss-Lobatto points more evenly in the bulk of the domain.

\begin{figure*}
\centering
 \includegraphics[width=0.95\textwidth]{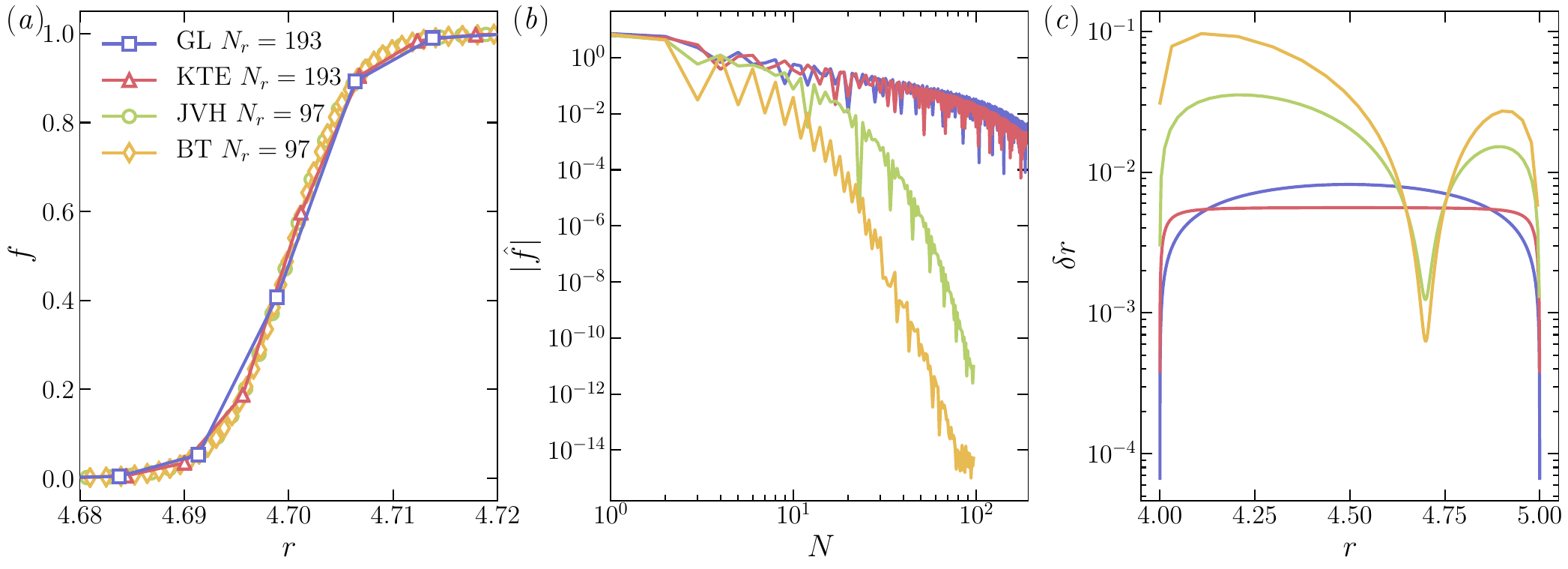}
 \caption{(\textit{a}) Example of a steep function 
$f(r)=\frac{1}{2}\left(1+\tanh \frac{r-\rmelt}{2\epsilon}\right)$ represented using the Gauss-Lobatto grid (GL), the KTE mapping by \citet{Kosloff93} with $\alpha_1=0.993$, the 
JVH mapping by \citet{Jafari15} and the BT mapping by \citet{Bayliss92} with $\alpha_1=40$ and $\alpha_2=0.4$ for 
$\epsilon=3\times 10^{-3}$ and $\rmelt=4.7$.
(\textit{b}) Absolute value of the coefficients of the discrete cosine transform of $f$ as a function of the degree of the Chebyshev polynomial $N$.
(\textit{c}) Grid spacing $\delta r$ as a function of $r$ for the same mappings.}
\label{fig:mappings}
\end{figure*}

As an illustrative example, Fig.~\ref{fig:mappings}(\textit{a}) shows the radial profile of a steep $\tanh$ function centered around the radius $\rmelt$ \citep[which is typically the kernel used in many phase field formulations, see][]{Beckermann99} using the GL and KTE grids
with $N_r=193$ and using the JVH and BT mappings with $N_r=97$, while Fig.~\ref{fig:mappings}(\textit{b}) shows the corresponding Chebyshev spectra as a function of the degree of the Chebyshev polynomial $N$.
To stress the differences in terms of grid resolution, Fig.~\ref{fig:mappings}(\textit{c}) shows the grid spacing $\delta r_k=|r_{k+1}-r_k|$ for $k\in[1,N_r-1]$  as a function of $r$.
GL and KTE are coarse with only $3$-$4$ collocation points 
distributed around the steep changes of the function. Using the BT or the JVH mappings allows to reduce the grid spacing by a factor five around the transition radius $\rmelt$ for half the number of radial grid points used in the GL or KTE mappings.
Comparison of the spectra in Fig.~\ref{fig:mappings}(\textit{b}) also shows that a much faster convergence of the collocation method towards machine precision is achieved for the former two mappings.
Of practical interest for fluid problems with boundary layers, we also note that the JVH mapping exhibits a slightly slower convergence than the BT mapping but retains a better resolution (i.e. a smaller grid spacing) at the spherical shell boundaries.
The chosen value of the interface thickness $\epsilon$ yields a 
constraint on the maximum eligible grid spacing \citep{Favier19}
\begin{equation}
 \max( \delta r) < 2\epsilon\,.
 \label{eq:reso}
\end{equation}
This constraint combined with the time-step restriction due to the 
volume-penalization \eqref{eq:courant_phase} makes the computation of phase field models much more numerically-demanding than the classical rotating convection problems. For $\epsilon\approx 10^{-3}-3\times 10^{-3}$ considered here, the radial resolution is about two to three times greater in phase field models than in
their monophasic counterparts to satisfy Eq.~\eqref{eq:reso}, while the time step size $\delta t$ is about a factor ten smaller to fulfill Eq.~\eqref{eq:courant_phase}.
This limits the range of physical parameters $E$ and $Ra$ accessible to global numerical modelling.

\subsection{Diagnostics}

The interface between the solid and liquid phases depends on the angular directions $(\vartheta,\varphi)$ and evolves in time. Following \cite{Favier19} and \cite{Yang23}, we choose to define it by
\begin{equation}
 \phase(r=\rmelt,\vartheta,\varphi,t)=\dfrac{1}{2}\,.
 \label{eq:rmelt_def}
\end{equation}
In the following, we employ angular brackets for spatial averaging
and overbars for time-averaging such that
\[
\begin{aligned}
 \langle f \rangle_V  & = \dfrac{1}{V}\int_V f\mathrm{d}V,
 &\langle f \rangle_S  & = \dfrac{1}{4\pi}  \int_0^{\pi} \int_0^{2\pi} f
\sin\vartheta\mathrm{d}\vartheta\mathrm{d}\varphi, \\
\langle f \rangle_\varphi & = \dfrac{1}{2\pi}\int_0^{2\pi} 
f\mathrm{d}\varphi, 
& \overline{f} & = \dfrac{1}{\tau_\text{avg}}\int_{t_0}^{t_0+\tau_\text{avg}} f 
\mathrm{d}t,
\end{aligned}
\]
where $t_0$ and $\tau_\text{avg}$ respectively denote the beginning and the width of the time-average interval, and $V$ is the spherical shell volume.
For clarity, we define in the following the time-averaged mean melt radius 
\begin{equation}
 \rmeltmean = \langle \overline{\rmelt} \rangle_S,
\end{equation}
and the corresponding thicknesses of the solid and liquid phases by 
\begin{equation}
\hS = r_o-\rmeltmean,\  \hL = \rmeltmean-r_i\,.
\end{equation}
The corresponding volumes are accordingly expressed by
\begin{equation}
 \volS = \int_0^{2\pi} \int_0^\pi 
\int_{\overline{\rmelt}(\vartheta,\varphi)}^{r_o} r^2 \sin\vartheta \mathrm{d}r 
\mathrm{d}\vartheta\mathrm{d}\varphi,\ \volL=V-\volS\,.
\end{equation}
For an easier comparison with standard monophasic convective models, we define effective quantities based on the actual mean thickness and temperature contrast of the fluid layer:
\begin{equation}
\label{eq:eff}
 \etaL=\dfrac{r_i}{\rmeltmean},\ \RaL= Ra \, g_m \Delta \tmelt \hL^3,\ 
\EL = \dfrac{E}{\hL^2}\,,
\end{equation}
where $\Delta \tmelt=1-\tmelt$ is the temperature 
contrast in the liquid and $g_m=\rmeltmean/r_o$ is the gravity at $\rmeltmean$.

The kinetic energy content per unit volume can be decomposed in poloidal and toroidal 
contribution following
\begin{equation}
\label{eq:ek}
 E_K = \dfrac{1}{2} \left( \langle \vec{u}_P^2 \rangle_V
 +  \langle \vec{u}_T^2 \rangle_V\right)
 = \sum_{\ell=1}^{\ell_\text{max}}
 E_{\ell}^P + \sum_{\ell=1}^{\ell_\text{max}} E_{\ell}^T\,,
\end{equation}
where $E_\ell^P$ ($E_\ell^T$) respectively denote the poloidal (toroidal) 
kinetic energy content at the spherical harmonic degree $\ell$.
The convective flow velocity is accordingly measured by a Reynolds number defined on the averaged liquid thickness $\hL$
\begin{equation}
Re_\mathcal{L} = Re\sqrt{\dfrac{V}{\volL}}\hL,\ 
 Re = \sqrt{2 \overline{E_K}}.
\end{equation}
Heat transfer is characterised by the Nusselt number here defined at the spherical shell boundaries
\begin{equation}
\label{eq:nu}
 Nu = \left.\eta\dfrac{\mathrm{d} \langle \overline{T}\rangle_S 
}{\mathrm{d}r}\right|_{r=r_i}
=\left.\dfrac{1}{\eta}\dfrac{\mathrm{d} \langle \overline{T}\rangle_S
}{\mathrm{d}r}\right|_{r=r_o}\,,
\end{equation}
where the factors involving the radius ratio $\eta$ come from the diffusive temperature gradient.
Again, for comparison purpose with standard models without phase change, it is insightful to also define an equivalent Nusselt number 
\begin{equation}
 \NuL = Nu \dfrac{\hL}{\Delta \tmelt} \dfrac{\etaL}{\eta}\,.
\end{equation}
Following \citet{Schwaiger19}, the typical convective flow lengthscale is defined as the time-average of the peaks of the poloidal kinetic energy spectra 
\begin{equation}
 \ell_U = \overline{\argmax_\ell E_\ell^P(t)}\,.
 \label{eq:ellU}
\end{equation}
To define a typical lengthscale for the topography of the 
solid-liquid interface, we proceed with a truncated spherical 
harmonic expansion such that
\begin{equation}
 \rmelt(\vartheta,\varphi,t) \approx 
\sum_{\ell=0}^{\ell_\text{max}}\sum_{m=-\ell}^\ell
\xi_{\ell m}(t) Y_{\ell m}(\vartheta,\varphi),
\label{eq:rmelt_sph}
\end{equation}
where $Y_{\ell m}$ denotes the spherical harmonic function of degree $\ell$ and order $m$.
Similarly to $\ell_U$, we then define $\ell_\xi$ as the time average of the peaks of the spectra defined by
\begin{equation}
 \ell_\xi = \overline{\argmax_\ell E_\ell^\xi(t)},\  E_\ell^\xi(t)= 
\sum_{m=-\ell ,m\neq0}^\ell |\xi_{\ell m}|^2\,.
 \label{eq:ellXi}
\end{equation}
Corresponding lengthscales at any radius read $\mathcal{L}_{[U,\xi]}(r) \approx 
\pi r/\ell_{[U,\xi]}$ \citep[e.g.][\S3.6.3]{Backus96}.

\subsection{Parameter coverage}


\begin{figure}
\centering
 \includegraphics[width=8.3cm]{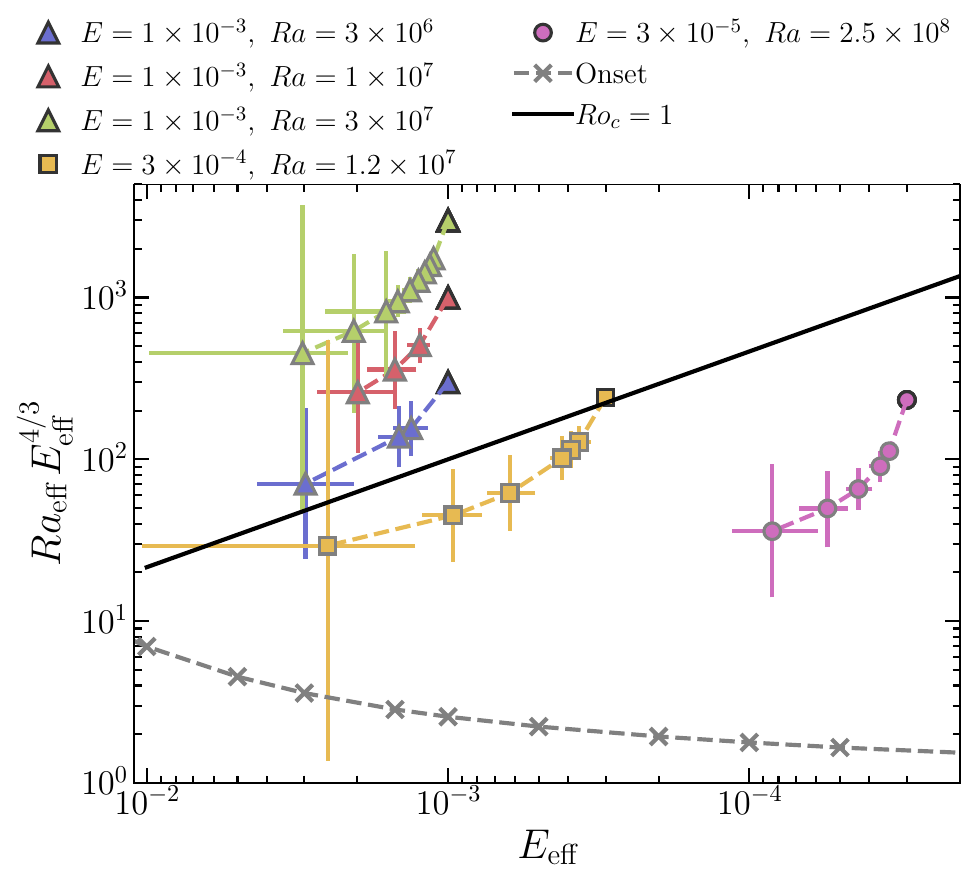}
 \caption{Numerical simulations carried out in this study located in a parameter space constructed using the effective Ekman and Rayleigh numbers $\EL$ and $\RaL$ defined in~\eqref{eq:eff}.
 Symbols with a black rim correspond to the configurations with no solid phase (i.e. $\tmelt=0$).
 Each type of symbols corresponds to fixed values of $Ra$ and $E$ and increasing values of $\tmelt$.
 The errorbars correspond to the control parameters constructed using $\min_{\vartheta,\varphi}(\overline{\rmelt})$ and 
$\max_{\vartheta,\varphi}(\overline{\rmelt})$.
They highlight the maximum topography of each model.
For comparison purpose, the critical Rayleigh numbers for onset of spherical shell convection with $r_i/r_o=0.8$ come from the study by \citet{Barik23}.
The solid black line corresponds to $Ro_c=1$ (Eq.~\ref{eq:roc}).}
 \label{fig:parameters}
\end{figure}

We carried out $26$ numerical simulations with $\eta=0.8$, $Pr=1$ and $St=1$ divided in four groups with fixed 
parameter pairs $(E,Ra)$ and increasing values of the melting temperature $\tmelt$.
The Stefan number $St$ defined by Eq.\eqref{eq:params} controls the time-scale separation between the temporal evolution of the phase-change interface and that of the flow.
While sub-glacial oceans are probably characterised by $St\gg1$ (using $\mathcal{L}\sim 3\times 10^5$~J/kg, $c_p=4\times 10^3$~J/kg/K from \cite{Journaux20} and a temperature gradient of $0.04$~K/km \citep{Vance18} yields $St\approx 40$ for Europa), we choose to consider the less numerically-demanding value $St=1$ which speeds up the melting and freezing dynamics leading to faster transients.
Recent studies on thermal convection interacting with a phase change boundary have shown that the value of the Stefan number has only a marginal impact on the final quasi-stationary equilibrium \citep{Rabba18,Purseed20,Yang23}.
Similarly, we choose to fix the value of the Prandtl number to $Pr=1$ for numerical convenience,
knowing that oceans are more likely to be characterised by $Pr\sim\mathcal{O}(10)$.

Spatial resolutions, control parameters and main diagnostics of these simulations are listed in 
Table~\ref{tab:results}.
A good indicator of the rotational constraint on the convective flow is
provided by the convective Rossby number 
\begin{equation}
 Ro_c = \sqrt{\dfrac{\RaL\EL^2}{Pr}}\,.
 \label{eq:roc}
\end{equation}
introduced by \citet{Gilman77}. Figure~\ref{fig:parameters} shows the location of the simulations in terms of their effective Ekman and Rayleigh numbers once the system has reached a quasi-stationary state.
Estimates of the transport properties in the subsurface oceans of the icy satellites yield Ekman and Rayleigh numbers that range from $E\sim 10^{-10}$ and $Ra\sim 10^{16}$ for smaller moons to $E\sim 10^{-13}$ and $Ra\sim 10^{24}$ for the larger ones \citep[for details, see e.g. Table~1 in][]{Soderlund19}, which is way outside the ranges considered
in Fig.~\ref{fig:parameters}.
We nonetheless see that our parameter choice yields configurations where $Ro_c$ exceeds unity  with $E=10^{-3}$ and $Ra\in[3\times 10^6, 10^7, 3\times 10^7]$ (triangles) and configurations where $Ro_c< 1$ with $E=3\times 10^{-4}$ and $Ra=1.2\times 10^7$ (squares) and $E=3\times 10^{-5}$ and $Ra=2.5\times 10^8$ (circles), a physical regime expected to be relevant to the subsurface oceans of the icy moons \citep[e.g.][their Fig.~1]{Soderlund19}.
For each group of simulations, an increase of $\tmelt$ goes along with a decrease of the effective  Rayleigh number $\RaL$ and an increase of the effective Ekman number $\EL$, as expected from the gradual decrease of the fluid layer thickness as $\tmelt$ increases.
The errorbars attached to each symbol correspond to the changes of the effective control parameters when considering the extrema 
of the interface $\min_{\vartheta,\varphi}(\overline{\rmelt})$ and 
$\max_{\vartheta,\varphi}(\overline{\rmelt})$ to evaluate them.
Overall, increasing $\tmelt$ at fixed 
$(E,Ra)$ yields a parameter path along which (\textit{i}) $Ro_c$ 
decreases, hence strengthening the rotational constraint; (\textit{ii}) the supercriticality of the convective flow drops; and (\textit{iii}) the topographic changes of the interface increase.

Most of the numerical simulations have been initiated from a diffusive state and a random noise temperature perturbation combined with a spherically-symmetric phase field $\tanh$ profile centered around $\tmelt$.
Simulations are run until a quasi-stationary state is reached at which point diagnostics are computed.
For the most demanding configurations, the simulations have been computed by restarting from setups with neighbouring parameters.
While bistability is known to occur in convective systems with phase change \citep{Purseed20,Yang23prl}, we have not observed such behaviour in our simulations.

\section{Results}
\label{sec:results}

\begin{figure*}
\centering
 \includegraphics[width=0.95\textwidth]{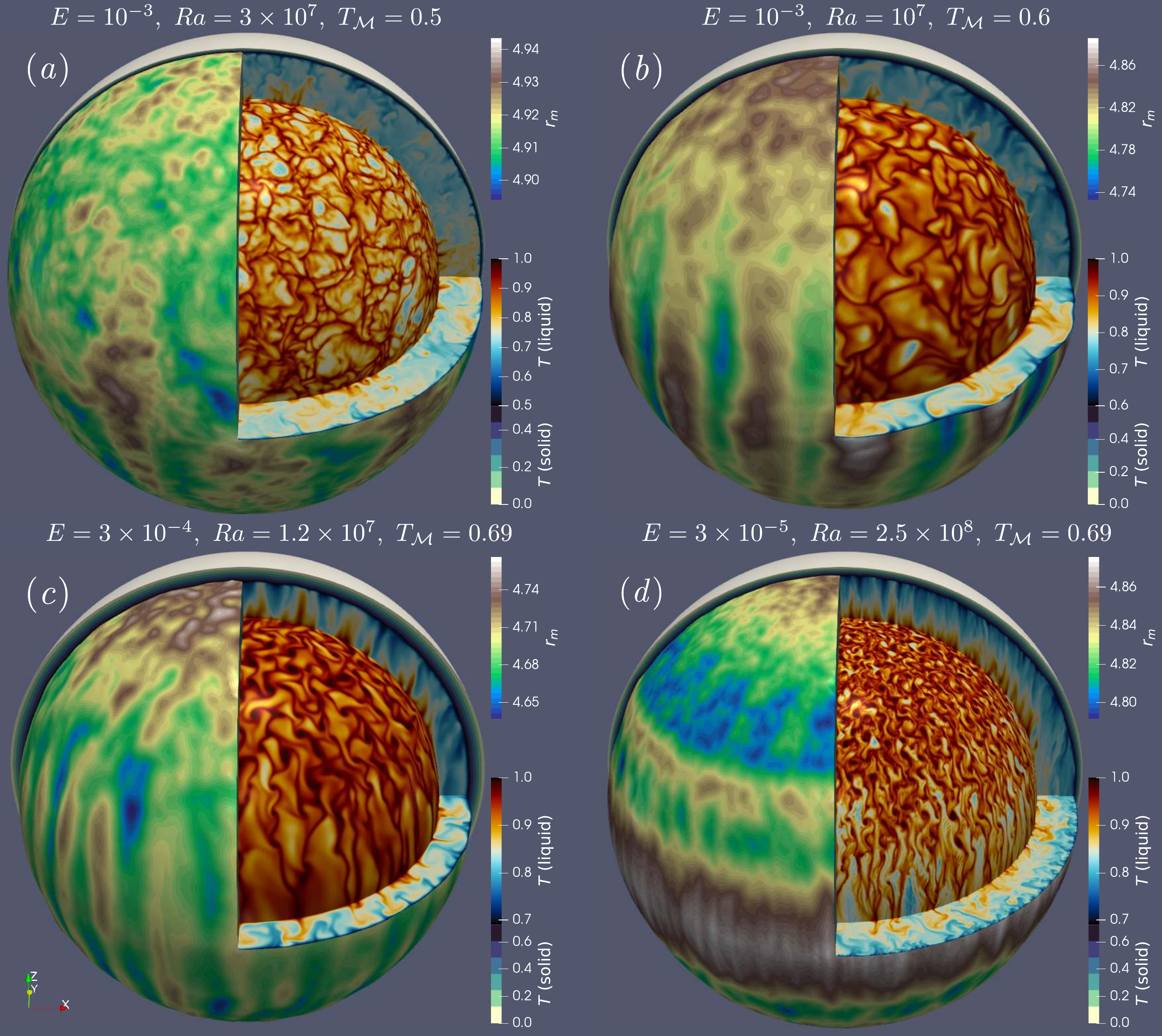}
\caption{Three-dimensional renderings of snapshots of four selected 
simulations. For each simulation, the inner sphere 
shows the temperature at $r=r_i+0.02$ atop the inner thermal boundary 
layer, while the outer surface corresponds to the melting radius $\rmelt$. 
Equatorial and meridional slices show the temperature in the liquid and solid 
phase with two different separated colormaps.}
\label{fig:snaps}
\end{figure*}

\begin{figure*}
\centering
  \includegraphics[width=0.95\textwidth]{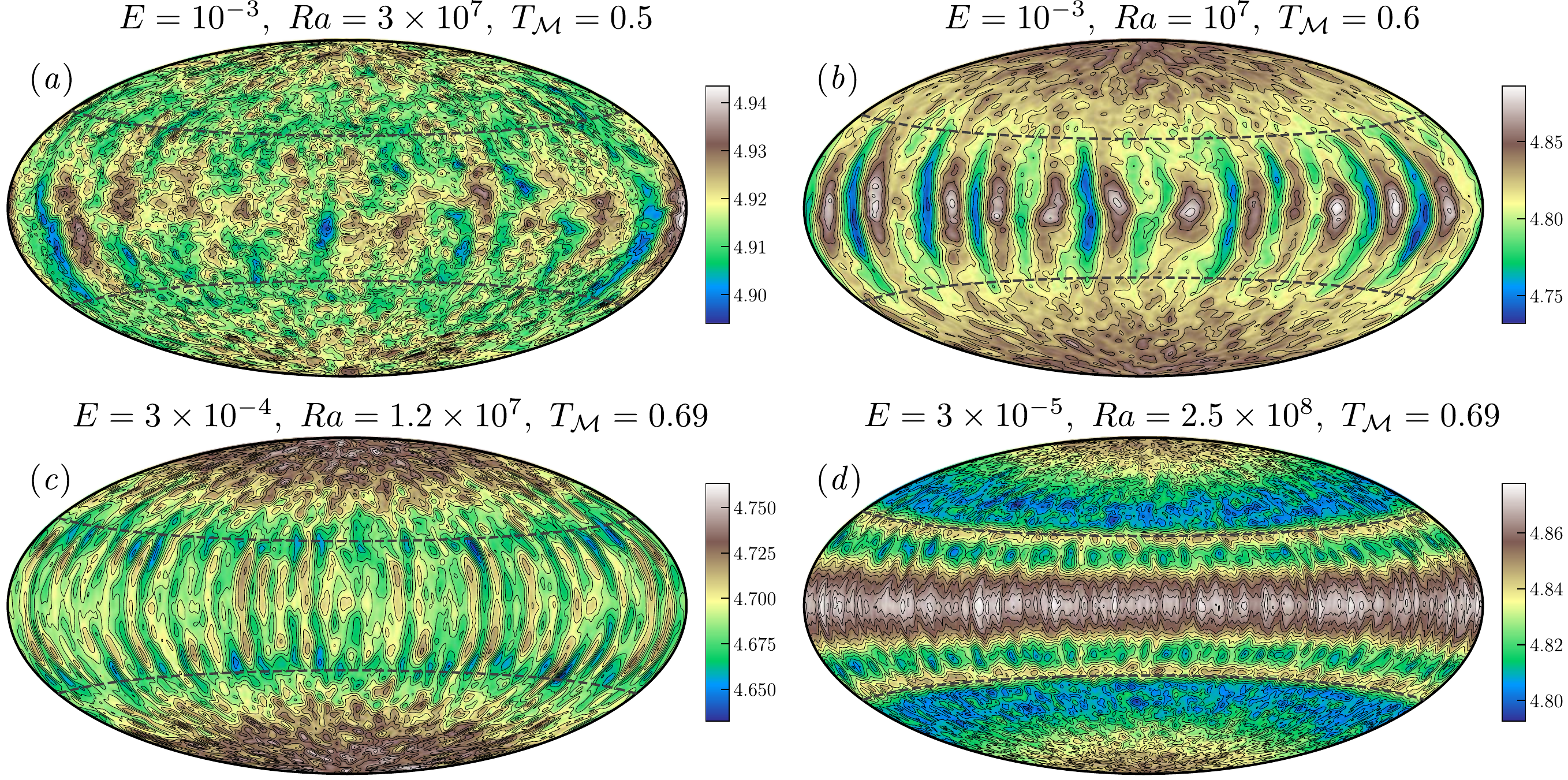}
 \caption{Hammer projections of snapshots of the melt radius 
$\rmelt$ for the four simulations shown in Fig.~\ref{fig:snaps}. In each 
panel, the dashed 
lines correspond to the location of the effective tangent cylinder, i.e.
$\vartheta=[\arcsin \etaL,\pi-\arcsin \etaL]$.}
 \label{fig:rmelts}
\end{figure*}

We first focus on selected simulations to highlight the interplay between the convective flow and the shape of the solid-liquid interface. 
Figure~\ref{fig:snaps} shows three-dimensional renderings of the temperature field for four selected simulations with an increasing rotational constraint on the flow, while Fig.~\ref{fig:rmelts} shows the corresponding Hammer projections of the melt radius $\rmelt$.

The first case with $E=10^{-3}$, $Ra=3\times 10^7$ and $\tmelt=0.5$ (Fig.~\ref{fig:snaps}\textit{a} and Fig.~\ref{fig:rmelts}\textit{a}) 
corresponds to weakly-rotating convection with $Ro_c\approx 4$. Similar to classical RBC, the thermal plumes are radially-oriented and the convective pattern at the edge of the inner thermal boundary layer forms a network of thin sheet-like upwellings surrounding broad downwellings. The solid-liquid interface is almost spherical, with little variations with $\rmelt\in[4.89,4.94]$ (we recall that we use $\eta=0.8$ so that $r\in[4,5]$) and no preferred direction, indicating the weak rotational constraint.
The second model with $E=10^{-3}$, $Ra=10^7$ and $\tmelt=0.6$ 
(Fig.~\ref{fig:snaps}\textit{b} and Fig.~\ref{fig:rmelts}\textit{b})
corresponds to $Ro_c \approx 2$. 
The convective pattern is more laminar than in the previous case due to a weaker effective Rayleigh number but remains similar regarding the influence of rotation. The most striking difference comes from the solid-liquid interface which now features large-scale topographic changes aligned with the rotation axis outside the effective tangent cylinder, defined by the colatitudes $\vartheta=[\arcsin \etaL,\pi-\arcsin \etaL ]$. 
Those features are reminiscent of the so-called ``banana cells'', i.e the convective columns outside the tangent cylinder of rotating spherical shells \citep[e.g.][]{Busse70,Simitev03}.
For that particular case, the amplitude of topography from trough to crest in the equatorial plane reaches about $15\%$ of the shell gap.
Similar to previous findings in non-rotating convection 
\citep[e.g.][]{Rabba18,Yang23}, cold plumes detach from the thermal boundary layer in the cusps of the solid-liquid interface (equatorial cut in Fig.~\ref{fig:snaps}\textit{b}). 
The influence of rotation is more pronounced in the third case  with $Ro_c\approx 0.7$ (Fig.~\ref{fig:snaps}\textit{c} and 
Fig.~\ref{fig:rmelts}\textit{c}).
This manifests itself by a stronger alignment of the thermal plume with the rotation axis inside the tangent cylinder (meridional cut in Fig.~\ref{fig:snaps}\textit{c}) and vortical structures at 
the connection points of the convective sheets (radial cut in 
Fig.~\ref{fig:snaps}\textit{c}). The solid-liquid interface is made up of two main contributions: (\textit{i})
an overall large-scale latitudinal contrast with thinner ice crust inside the effective tangent cylinder than outside; (\textit{ii}) small-scale columnar corrugations aligned with the rotation axis outside the tangent cylinder.
The decrease of the effective Ekman number compared to the previous case goes along with smaller scale columnar convection and weaker topographic changes in the equatorial region.
This trend is confirmed in the last simulation with the smallest Ekman number and $Ro_c\approx 0.3$
(Fig.~\ref{fig:snaps}\textit{d} and Fig.~\ref{fig:rmelts}\textit{d}).
In this rotationnally-constrained configuration, most of the topographic changes are axisymmetric with a thinner ice thickness at the equator than at the poles. 
The melt radius also shows secondary peaks right at the location of the tangent cylinder, a feature reminiscent of the local flux maxima observed there in rotating spherical shell convection \citep[see Fig.~3 in][]{Gastine23}.

Overall the increasing influence of rotation goes along with a topography which transits from columnar troughs and crests outside the effective tangent cylinder to an almost axisymmetric profile modulated by small-scale roughness.
This prompts us to separately investigate the large-scale 
axisymmetric topography and the non-axisymmetric features.
In the following, we hence disentangle the axisymmetric topography defined by
\begin{equation}
 \tilde{\xi} = \langle \overline{\rmelt} \rangle_\varphi\,,
\label{eq:mean_field_topo}
\end{equation}
from the fluctuating non-axisymmetric patterns defined by the following standard
deviation
\begin{equation}
\xi'=\overline{\left(\langle r_\mathcal{M}^2 \rangle_\varphi - \langle \rmelt 
\rangle_\varphi^2\right)^{1/2}}
\,.
\label{eq:fluct_field_topo}
\end{equation}

\subsection{Axisymmetric topography}

\begin{figure*}
 \centering
 \includegraphics[width=0.95\textwidth]{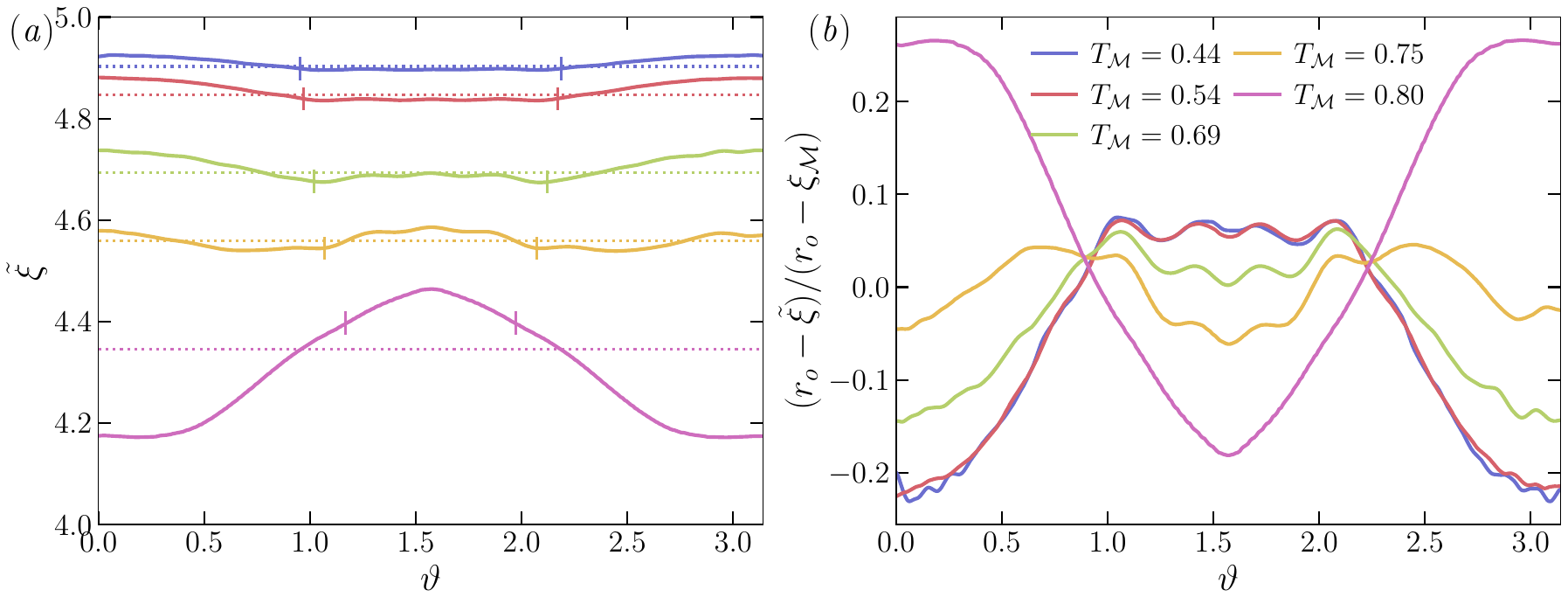}
 \caption{(\textit{a}) Time and azimuthal average of the melt radius 
$\tilde{\xi}$ (see Eq.~\ref{eq:mean_field_topo}) as a function of colatitude 
for a series of numerical simulations with $E=3\times 10^{-4}$ and $Ra=1.2\times 
10^7$ and increasing $\tmelt$. (\textit{b}) Normalised thickness of the solid 
phase $(r_o-\tilde{\xi})/(r_o-\rmeltmean)$ as a function of colatitude. 
The dotted lines in panel (\textit{a}) correspond to the mean melt radius 
$\rmeltmean$, 
 while the vertical segments mark the location of the effective 
tangent cylinder.}
\label{fig:theta}
\end{figure*}

Figure~\ref{fig:theta}(\textit{a}) shows latitudinal profiles of the mean axisymmetric melt radius $\tilde{\xi}$ for a series of simulations with $E=3\times 10^{-4}$ and $Ra=1.2\times 10^7$.
An increase in $\tmelt$ goes along with thicker ice with 
increasing latitudinal contrasts.
For an easier comparison,  Fig.~\ref{fig:theta}(\textit{b}) shows the corresponding latitudinal variations of the ice thickness normalised by its mean value, i.e. $(r_o-\tilde{\xi})/(r_o-\rmeltmean)$.
Relative changes in the ice thickness are found to be quite similar from one simulation to another, ranging between $-0.2$ and $0.2$.
The axisymmetric topography evolves from configurations with thicker ice at the equator than at the poles for $\tmelt < 
0.7$ to the opposite for $\tmelt > 0.75$.
Postponing for now the question of the control parameters which govern this transition, we note a good correlation between the 
profiles shown in Fig.~\ref{fig:theta}(\textit{b}) and the outer boundary heat flux obtained in the monophasic rotating convection models by \citet{Amit20} (their Fig.~7) and by \citet{Kvorka22} (their Fig.~10).

\begin{figure*}
 \centering
  \includegraphics[width=0.95\textwidth]{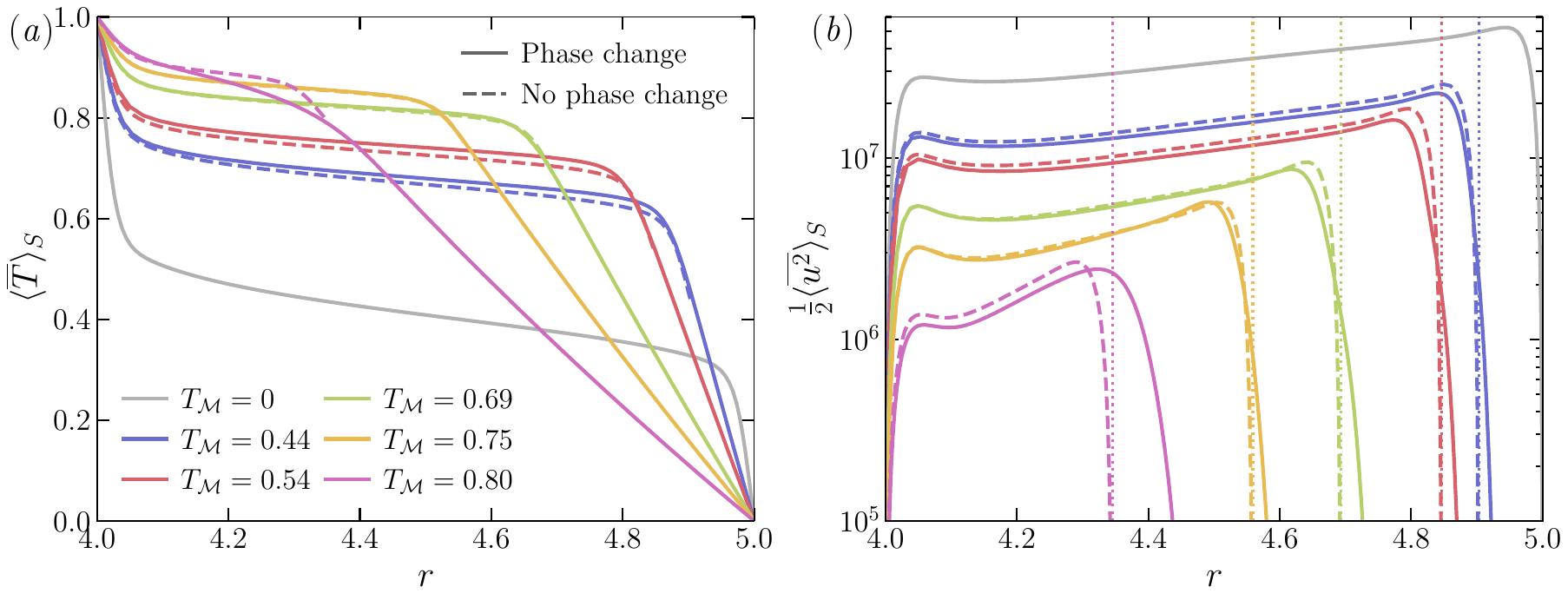}
  \caption{Comparison between models with a phase change and their equivalent convective models 
without for simulations with $E=3\times 10^{-4}$ and $Ra=1.2\times 10^7$. (\textit{a}) Time-
averaged radial profiles of temperature for increasing values of the melting temperature $\tmelt$.
(\textit{b}) Time-averaged radial profiles of kinetic energy. For comparison purpose, the grey lines in panels (\textit{a}) and (\textit{b}) correspond to the setup with no solid phase (i.e. $\tmelt=0$). The dotted vertical lines in panel (\textit{b}) correspond to $\rmeltmean$.}
\label{fig:equiv}
\includegraphics[width=.95\textwidth]{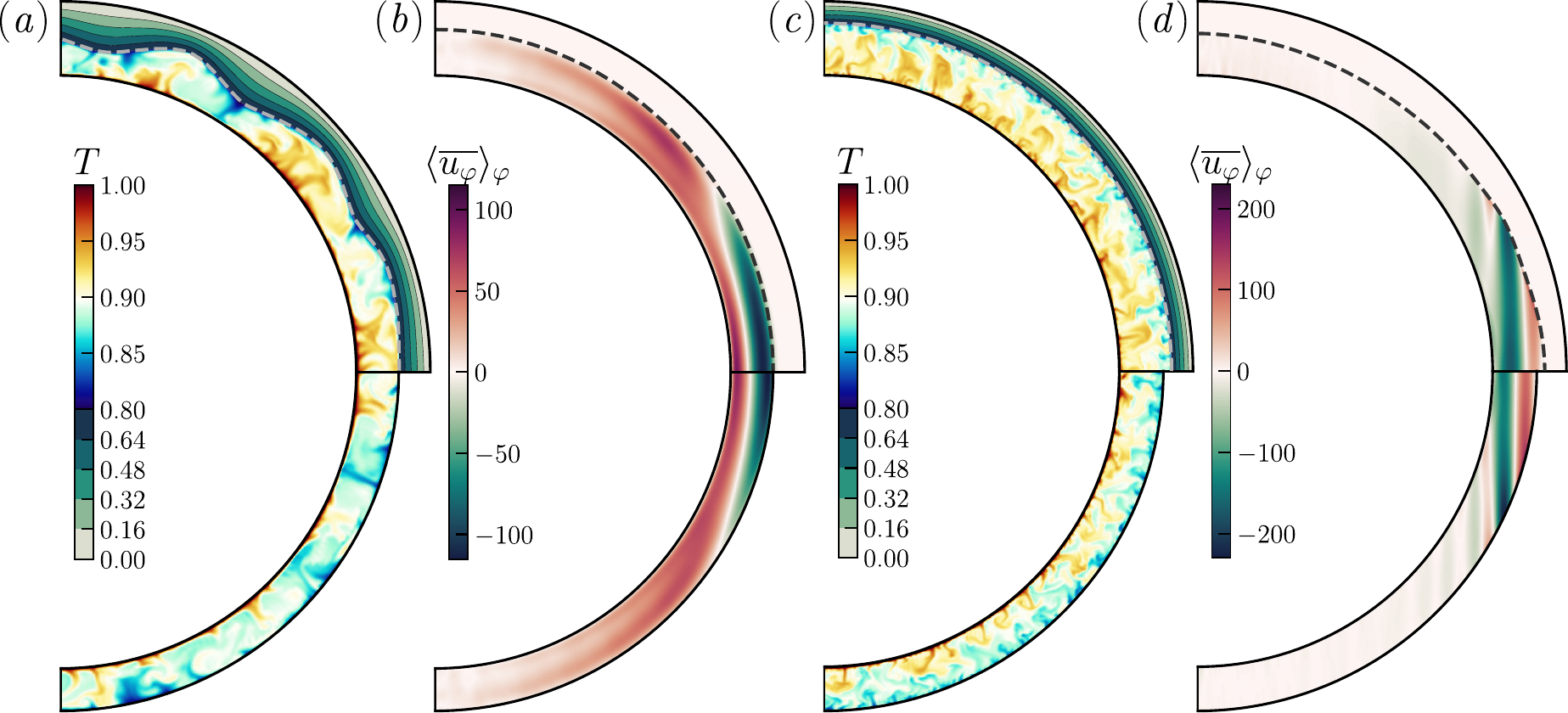}
\caption{Comparison between models with a phase change (upper half of each panel) with equivalent 
convective models without (lower halves). Panels (\textit{a}) and (\textit{c}) show snapshots of the 
temperature in the equatorial plane, while panels (\textit{b}) and (\textit{d}) show the time and 
azimuthal average of $u_\varphi$ in a meridional plane. Panels (\textit{a}) and (\textit{b}) 
correspond to a weakly-rotating configuration with $E=10^{-3}$, $Ra=3\times 10^7$, $\tmelt=0.8$, 
while (\textit{c}) and (\textit{d}) correspond to a rotationnally-constrained setup with  $E=3\times 
10^{-5}$, $Ra=3\times 10^8$, $\tmelt=0.8$. The dashed lines mark the location of the solid-liquid 
interface.}
\label{fig:comp_hydro}
\end{figure*}

This raises the question of the feedback of the topography onto the flow.
To examine this issue, we have computed the equivalent $26$ rotating spherical shell simulations without phase change adopting the effective control parameters $\RaL$, $\EL$ and $\etaL$.
In addition, we have also carried out $5$ extra simulations without solid phase (i.e. $\tmelt=0$) for each $(Ra,E)$ pair.
Figure~\ref{fig:equiv} shows a comparison between the mean radial profiles of temperature (panel \textit{a}) and kinetic energy (panel \textit{b})for numerical models with a phase change (solid lines) and their equivalent counterparts without (dashed lines).
Typical of rotating convection, the temperature profile of the reference case with no solid phase ($\tmelt=0$) is made up of three distinct parts: two thermal boundary layers which accommodate most of the temperature contrast and a quasi-linear temperature drop in the fluid bulk.
The corresponding kinetic energy profile features two localised maxima which mark the location of the edges of the Ekman boundary layers.
The increase of $\tmelt$ goes along with a gradual decrease of the kinetic energy content due to the shrinking of the fluid region.
Temperature then follows a quasi-linear conducting profile in the solid part.
Models with or without phase changes almost perfectly overlap for $\tmelt \leq 0.75$.
They depart more strongly for the largest melting temperature $\tmelt = 0.8$, due to the largest amplitude of topography in this configuration (recall Fig.~\ref{fig:theta}\textit{a}).
In view of the good agreement between the radial profiles, 
the average rms properties of the convective flow are found to be little affected by the phase change (see Table~\ref{tab:results}).

Figure~\ref{fig:comp_hydro} shows an additional comparison between
models with and without phase change for two configurations with large melting temperature to enhance the topographic changes. Panels (\textit{a}) and (\textit{b}) correspond to a 
weakly-rotating configuration with $E=10^{-3}$, $Ra=3\times 10^7$, $\tmelt=0.8$ and $Ro_c\approx 3.1$.
In this setup, the solid-liquid interface presents significant changes in the azimuthal direction.
Cold plumes which detach from the upper boundary layer are mostly localised in the cusps of the interface, while the hot upwellings are clustered in fluid regions with a thinner ice crust.
Despite those topographic changes, the mean azimuthal zonal 
flows $\langle \overline{u_\varphi} \rangle_\varphi$ appear strikingly similar between the phase field case and its monophasic counterpart. Equatorial zonal flows are retrograde close to the outer boundary, a typical feature of spherical-shell rotating convection when the convective Rossby number exceeds one \citep[e.g][]{Gilman77,Aurnou07,Gastine13,Yadav16}.
Zonal flow  gradients are predominantly radial, with no marked alignment with the axis of rotation, indicating a weak rotational constraint. Their energetic content amounts to about $4\%$ of the total kinetic energy in
that case.
Conversely, Fig.~\ref{fig:comp_hydro}(\textit{c}) and (\textit{d}) correspond to a rapidly-rotating configuration with $E=3\times 10^{-5}$, $Ra=2.5\times 10^8$, $\tmelt=0.8$ which yields $Ro_c \approx 0.3$. In this setup, topographic changes happen mostly in the latitudinal direction with a thinner ice at the equator than at the poles. 
As such, the fluid domain in the equatorial plane is therefore slightly thicker than its equivalent without phase change  
(Fig.~\ref{fig:comp_hydro}\textit{c}).
The typical size and shape of the convective plumes nevertheless show a good agreement between the two simulations.
Zonal flow profiles (Fig.~\ref{fig:comp_hydro}\textit{d}) take the 
form of a pair of alternated retrograde and prograde jets mostly localised outside the tangent cylinder, very much alike those obtained in monophasic spherical shell convection with rigid boundaries and comparable convective Rossby numbers $Ro_c \approx 0.3$ \citep[e.g.][rightmost simulation in their Fig.~5]{Cabanes24}. Their energetic content remains however quite weak, about $3\%$ of the total kinetic energy, a value in line with those reported by \citet{Cabanes24} for similar effective Ekman numbers 
$\EL\approx 10^{-4}$ (see their Table~1).
Despite the sizeable axisymmetric topographic variations, zonal flows appear little affected by the phase change. Assessing the feedback between topography and zonal jets would require reaching lower (larger) Ekman (Rayleigh) numbers such that they represent a greater share of the kinetic energy.

\begin{figure}
 \centering
 \includegraphics[width=.49\textwidth]{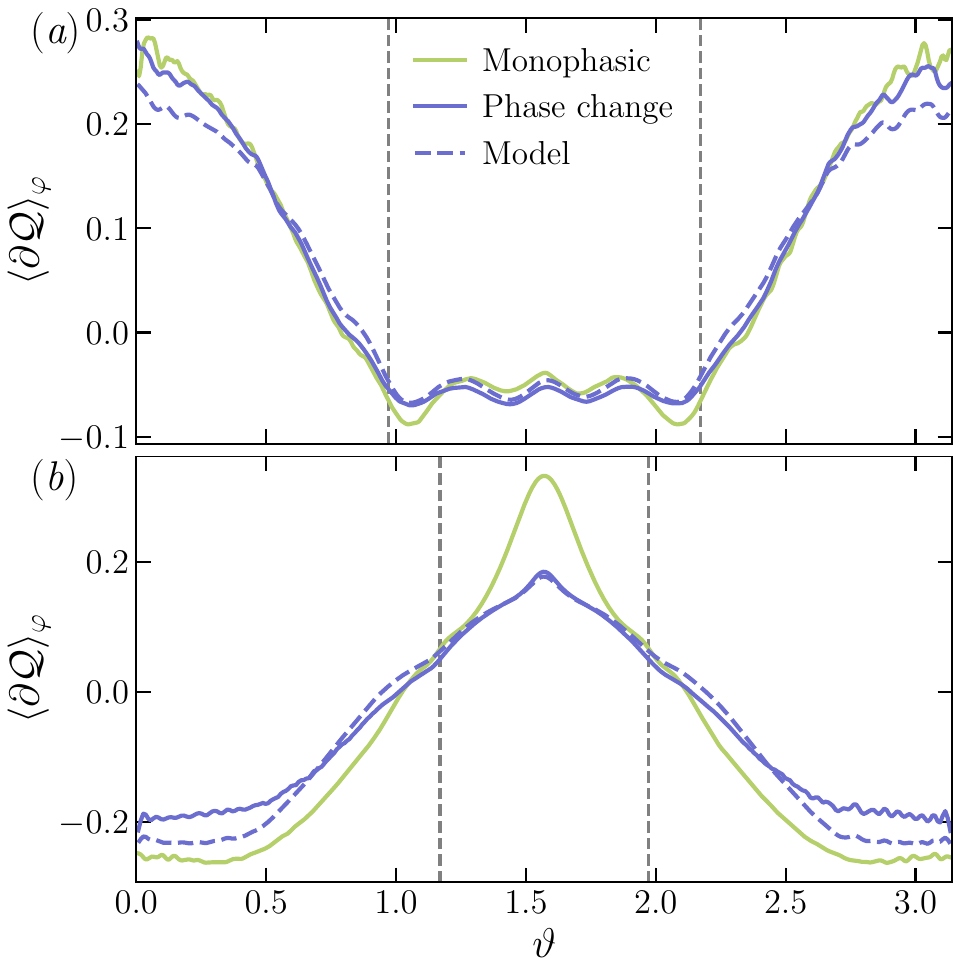}
 \caption{(\textit{a}) Time and azimuthal average of the relative heat flux 
\eqref{eq:flux_rel} expressed at 
the top of the liquid phase for a numerical simulation with $E=3\times 
10^{-4}$, $Ra=1.2\times 10^7$ and $\tmelt=0.55$. (\textit{b}) Same quantities 
for a numerical simulation with $E=3\times 10^{-4}$, $Ra=1.2\times 10^7$ and 
$\tmelt=0.7$. In both panels, the 
dashed vertical lines correspond to the location of the tangent cylinder.}
\label{fig:comp_heatf_hsolid}
\end{figure}

We finally examine the differences in terms of heat flux at the top of the 
fluid layer. To do so, we derive in Appendix~\S~\ref{sec:app_topo_theo} a perturbative model for the diffusion of heat in a solid phase with an upper spherical 
boundary held at a constant temperature $T=0$ and a lower quasi-spherical boundary with topographic changes held at $T=\tmelt$. 
Provided the amplitude of topography 
remains small compared to the radius, i.e. the spherical harmonic expansion 
coefficients from Eq.~\eqref{eq:rmelt_sph} fulfill $|\overline{\xi_{\ell m}}| / 
\rmeltmean \ll 1$, we show in Appendix~\ref{sec:app_topo_theo}
that the temperature gradient along the interface reads

\begin{equation}
 \partial \mathcal{Q}(\vartheta,\varphi) \approx \sum_{\ell\neq 0, m} 
\dfrac{\overline{\xi_{\ell m}}}{\rmeltmean}
f_\ell(\eta_S)
Y_{\ell m}(\vartheta,\varphi)\,,
\label{eq:topo_theo}
\end{equation}
with
\begin{equation}
 f_\ell(\eta_S)=\dfrac{\ell-1+(\ell+2)\eta_S^{2\ell+1}}{1-\eta_S^{2\ell+1}},
\end{equation}
and $\eta_S=\rmeltmean/r_o$ is the mean radius ratio of the solid phase and $\partial \mathcal{Q}$ characterises the relative heat flux changes
\begin{equation}
 \partial \mathcal{Q}(\rmelt,\vartheta,\varphi) = \dfrac{\dfrac{\partial 
\overline{T}}{\partial 
r}(\rmelt,\vartheta,\varphi)-\dfrac{\mathrm{d} T_0}{\mathrm{d r}}({ 
\rmeltmean})}{%
\dfrac{\mathrm{d} T_0}{\mathrm{d r}}(\rmeltmean)},
\label{eq:flux_rel}
\end{equation}
with 

\begin{equation}
 \dfrac{\mathrm{d}T_0}{\mathrm{d}r}=-\dfrac{\tmelt}{\hS\eta_S},
\end{equation}
the zeroth order diffusive temperature gradient in a spherical shell of mean 
inner radius $\rmeltmean$ and outer radius $r_o$.

To evaluate this first-order model, Fig.~\ref{fig:comp_heatf_hsolid} shows  a 
comparison of the relative variations of the axisymmetric heat flux $\langle 
\partial \mathcal{Q} \rangle_\varphi$  for  
two numerical simulations with $E=3\times 10^{-4}$ and $Ra=1.2\times 10^7$
which only differ from their melting temperature with $\tmelt=0.54$ and a thin 
ice thickness $\hS=0.15$ (panel \textit{a}) and $\tmelt=0.8$ and a thick ice 
shell $\hS=0.65$ (panel \textit{b}). 
To also examine the influence of the topography on the flow, we 
include in Fig.~\ref{fig:comp_heatf_hsolid} the heat flux profiles 
of the equivalent models without phase change. In that case, the relative heat 
flux variations are expressed by
\begin{equation}
 \langle \partial \mathcal{Q} \rangle_\varphi = \dfrac{\dfrac{\partial \langle 
\overline{T}\rangle_\varphi}{\partial r}(r_o,\vartheta)-
\dfrac{\mathrm{d} \langle 
\overline{T}\rangle_S}{\mathrm{d} r}(r_o)}{%
\dfrac{\mathrm{d} \langle 
\overline{T}\rangle_S}{\mathrm{d} r}(r_o)}\,.
\label{eq:flux_rel_hydro}
\end{equation}
The heat flux profiles shown in Fig.~\ref{fig:comp_heatf_hsolid}(\textit{a}) look very much alike those obtained 
by \citet{Amit20} in their so-called ``polar-cooling'' configurations (see the third panel in their Fig.~7).
In this setup, the heat flux is almost constant outside the tangent cylinder and increases gradually inside to 
reach local maxima in the polar regions.
We observe, on the one hand, an excellent agreement between the actual heat flux obtained in the phase field 
simulation and the theoretical model expressed in Eq.~\eqref{eq:flux_rel}, and on the other hand, very similar 
profiles with or without phase change. This latter observation indicates a negligible influence of the 
axisymmetric topography on the heat flux, which comes to no surprise given the small variations of the ice shell 
thickness in this configuration ($\tilde{\xi}\in[4.84,4.88]$, see 
Fig.~\ref{fig:theta}\textit{a}). Figure~\ref{fig:comp_heatf_hsolid}(\textit{b}) corresponds to a configuration with a smaller convective Rossby number $Ro_c\approx 0.67$. In that case, and in line with previous findings of monophasic
rotating convection \citep[see, e.g. the first panel of Fig.~7 in][]{Amit20}, the heat flux peaks in the equatorial region, a hallmark of the so-called ``equatorial cooling regime''. More pronounced differences between the monophasic and the phase field configurations are however observed:
while the perturbative model still correctly accounts for the actual heat flux, the configuration without phase changes now yields larger latitudinal heat variations than observed in the case with a phase change.
This clearly indicates than the purely spherical analogue cannot account for the heat flux changes for configurations with large topographic variations (here $\tilde{\xi}\in[4.17,4.46]$, see Fig.~\ref{fig:theta}\textit{a}).

\subsection{Non-axisymmetric roughness}

\begin{figure*}
 \centering
 \includegraphics[width=0.95\textwidth]{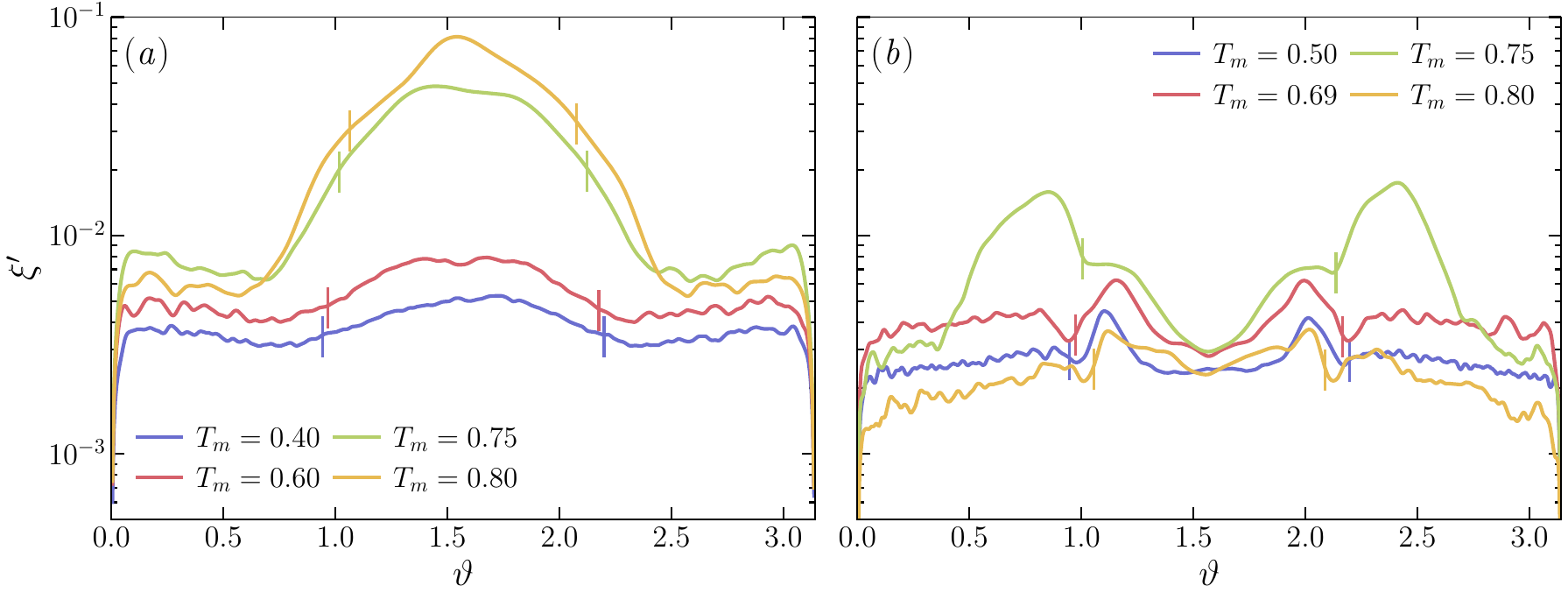}
 \caption{Mean latitudinal profile of non-axisymmetric topography (as defined 
in Eq.~\ref{eq:fluct_field_topo}) for two series of 
simulations with $E=10^{-3}$, $Ra=3\times 10^7$ (panel \textit{a}), and 
with $E=3\times 10^{-5}$, $Ra=2.5\times 10^8$ (panel \textit{b}). In each 
panel, the vertical segments mark the location of the tangent cylinder.}
\label{fig:comp_roughness}
\end{figure*}

We now turn our attention to the non-axisymmetric topography, characterised in 
terms of the standard deviation of $\rmelt$ along the longitudinal direction 
(Eq.~\ref{eq:fluct_field_topo}).

We have already seen in Fig.~\ref{fig:rmelts} significant regionalized differences in the amplitude and size of the interface roughness depending on the strength of the rotational constraint.
To illustrate this phenomenon, Fig.~\ref{fig:comp_roughness} shows mean latitudinal contrasts of $\xi'$ for two series of simulations with increasing melting temperature and with either a weak rotational constraint ($E=10^{-3}$, $Ra=3\times 10^7$, panel \textit{a}) or a strong influence of rotation ($E=3\times 10^{-5}$, $Ra=2.5\times 10^8$, panel \textit{b}).
For the former series of simulations, an increase of $\tmelt$ goes along with a gradual increase of the interface roughness.
The configurations which are the least influenced by rotation, i.e. $\tmelt \leq 0.6$ and $Ro_c >3.5$, feature almost no latitudinal variation and $\xi'$ does not exceed 1\% of the shell gap.
When $\tmelt  \geq 0.75$, the amplitude of the non-axisymmetric topography raises by one order of magnitude in the low-latitude 
regions reaching $\max_\vartheta \xi'\approx 0.1$.
This corresponds to the columnar topographic changes visible in Fig.~\ref{fig:rmelts}(\textit{b}).
The second series of simulations features more undulating profiles of overall weaker amplitude (Fig.~\ref{fig:comp_roughness}\textit{b}).
The amplitude of $\xi'$ follows a non-monotonic behaviour with the increase of $\tmelt$.
Latitudinal changes appear to be correlated with the variations of the mean axisymmetric ice thickness with localised minima at the equator and at the location of the tangent cylinder (see Fig.~\ref{fig:snaps}\textit{d}).

\begin{figure}
 \centering
  \includegraphics[width=0.49\textwidth]{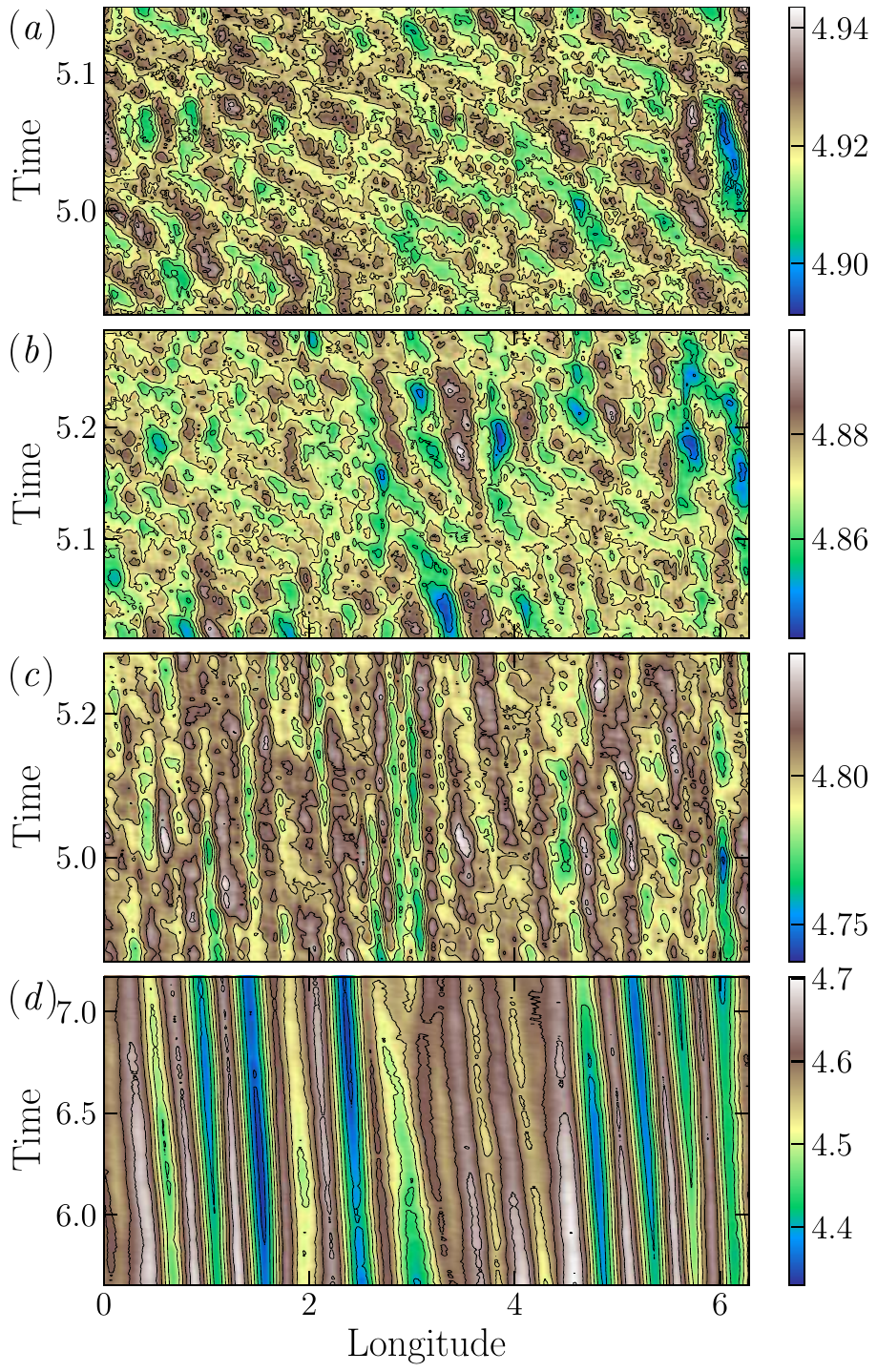}
 \caption{Longitudinal Hovmöller diagrams for the melt radius in the equatorial 
plane $\rmelt(\vartheta=\pi/2,\varphi,t)$ for four numerical simulations with 
$E=10^{-3}$ and $Ra=3\times 10^7$ with $\tmelt=0.5$ (panel \textit{a}), 
$\tmelt=0.6$ (panel \textit{b}), $\tmelt=0.69$ (panel \textit{c}) and 
$\tmelt=0.8$ (panel \textit{d}).}
 \label{fig:drift}
\end{figure}

To explore the time variability of the solid-liquid interface, we show in Fig.~\ref{fig:drift} Hovmöller diagrams of the melt radius in the equatorial plane $\rmelt(\pi/2,\varphi,t)$ for four simulations with $E=10^{-3}$, $Ra=3\times 10^7$ and increasing values of $\tmelt$.
The first two cases ($\tmelt=0.5$ and $\tmelt=0.6$) present rapid temporal variability on time scales of less than one tenth of the viscous diffusion time. Topography in the equatorial plane mostly drifts westward and the crest-to-trough amplitude is about $2$-$3$\% of the spherical shell gap.
In the second case with $\tmelt=0.6$, 
larger longer-lived topographic features with a larger amplitude are occasionally observed (e.g. $t\approx 5.2$ and $\varphi \approx 3.5$).
This trend becomes more pronounced in the third and fourth cases with $\tmelt = 0.69$ and $\tmelt =0.8$.
Topographic contrast increases and remains stable and coherent over timespans that exceed the viscous diffusion time. 
We only observe a very slow westward drift of the topographic features.
These latter two configurations are reminiscent of rotating convection models in Cartesian geometry by \citet{Ravichandran21} in which convective features are locked within the interface topography.
In contrast with the moderate feedback of the axisymmetric topography on the convective flow discussed before, non-axisymmetric roughness hence yields significant differences with 
classical rotating convection without phase change.
We recall that the Stefan number $St$ is here fixed to unity across all simulations.
While systematically varying $St$ is outside the scope of the current study, it would influence the characteristic timescale of topographic features with larger Stefan numbers leading to slower topography dynamics.

\begin{figure*}
 \centering
 \includegraphics[width=.95\textwidth]{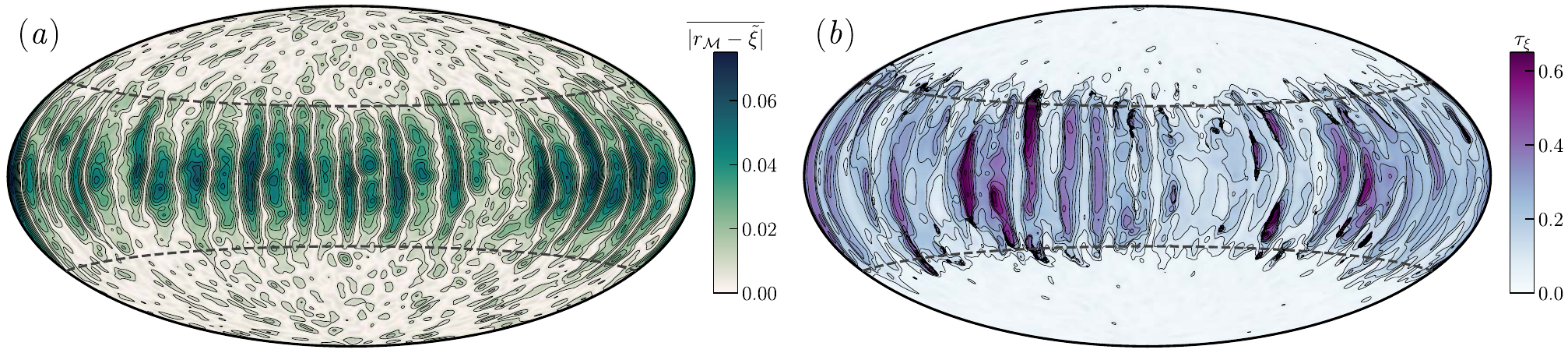}
 \caption{(\textit{a}) Hammer projection of the time-averaged non-axisymmetric topography defined by $|\overline{\rmelt-\tilde{\xi}}|$ for a numerical model with $E=10^{-3}$, $Ra=10^7$ and $\tmelt=0.6$. 
 Time averaging has been conducted over one thermal diffusion time of the ice layer, i.e. $Pr\hL^2\approx 3.4\times10^{-2}$.
 (\textit{b}) Corresponding Hammer 
projection of the correlation time of the topography expressed by 
Eq.~\eqref{eq:tau_topo}. In each panel, the dashed lines mark the location of the effective tangent cylinder.}
\label{fig:rmelt_tau}
\end{figure*}

To quantify the typical time of variation of the topography, we define
the following auto-correlation function for each location  
$(\vartheta,\varphi)$:
\[
\mathcal{C}(\vartheta,\varphi,\tau)=
\dfrac{\overline{\rmelt(\vartheta,\varphi,t+\tau)\rmelt(\vartheta,\varphi,t)}}
{\overline{\rmelt^2(\vartheta,\varphi,t)}}\,.
\]
The correlation time of the topography at each point 
$\tautopo(\vartheta,\varphi)$ is then defined as the 
full width at half maximum of $\mathcal{C}$, i.e.
\begin{equation}
\mathcal{C}(\vartheta,\varphi,\tautopo)=\dfrac{1}{2} \,. 
\label{eq:tau_topo}
\end{equation}
Figure~\ref{fig:rmelt_tau} shows an illustration of the time-averaged non-axisymmetric topography 
(panel \textit{a}) alongside the correlation time $\tautopo(\vartheta,\varphi)$ for a model with $E=10^{-3}$, $Ra=10^7$ and $\tmelt=0.6$. Given that the topography slowly evolves in that case, the time-averaging involved in panel (\textit{a}) has been conducted over the typical thermal diffusion time of the ice crust $Pr\hS^2\approx3.4 \times 10^{-2}$, a value that is about one order of magnitude below the typical time of variation of the topography in that case.
For both diagnostics, important regionalized differences are 
clearly visible between flow regions inside and outside the tangent cylinder. 
Most of the topographic changes, associated with the large scale columnar troughs and crests already seen in Fig.~\ref{fig:rmelts}(\textit{b}), indeed reside outside the tangent cylinder.
This large-scale non-axisymmetric topography is correlated with a corresponding increase of the typical time $\tautopo$ which locally exceeds a few tenths of the viscous diffusion time.

\begin{figure}
 \centering
 \includegraphics[width=.49\textwidth]{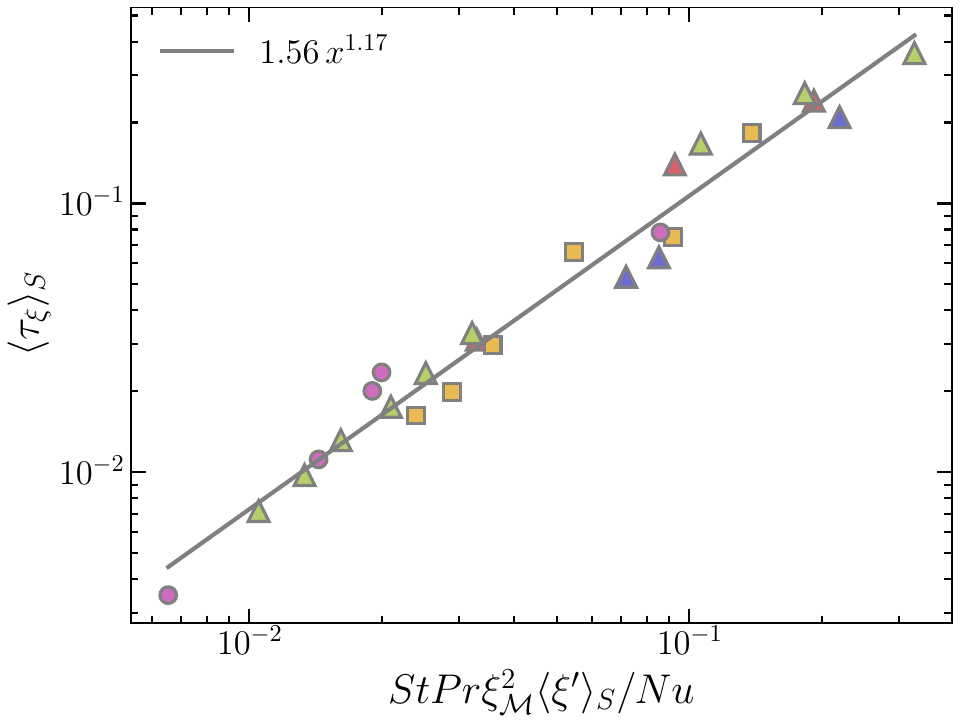}
 \caption{Average of the correlation time of the topography 
(Eq.~\ref{eq:tau_topo}) as a function of the theoretical 
scaling \eqref{eq:tautopo_scaling}. 
The solid line corresponds to a polynomial fit to the data. 
Symbols convey the same meaning as in Fig.~\ref{fig:parameters}.}
\label{fig:tau_topo}
\end{figure}

To further analyse how $\tautopo$ relates to the roughness amplitude, we define
\[
 \tilde{\mathcal{Q}} = -\dfrac{\partial \langle \overline{T} \rangle_\varphi 
}{\partial r}, \ \mathcal{Q}'= \overline{ \left[\left\langle\left(\dfrac{\partial 
T}{\partial r}\right)^2\right\rangle_\varphi-\left\langle \dfrac{\partial T}{\partial 
r}\right\rangle_\varphi^2\right]^{1/2}},
\]
following the decomposition already adopted in Eqs.~\eqref{eq:mean_field_topo}-\eqref{eq:fluct_field_topo}.
An horizontal and time average of Stefan's condition \eqref{eq:stefan} yields
\begin{equation}
 St Pr \dfrac{\mathrm{d} \rmeltmean}{\mathrm{d} t} = 0 = 
 \langle \tilde{\mathcal{Q}} \rangle_S -\dfrac{\tmelt 
r_o}{(r_o-\rmeltmean)\rmeltmean},
\end{equation}
where $\langle \tilde{\mathcal{Q}} \rangle_S = Nu\,r_i r_o/\rmeltmean^2$. 
The typical time variability of the interface roughness hence relates to the heat flux fluctuations, such that
\[
 St Pr  \dfrac{\mathrm{d} \langle \xi' \rangle_S}{\mathrm{d} t} \approx
 \langle \mathcal{Q}' \rangle_S\,.
\]
Following \cite{Yang23}, we make the additional hypothesis that the amplitude of the heat flux fluctuations are proportional to the average heat flux, i.e. 
$ \langle \mathcal{Q}' \rangle_S \sim  \langle \tilde{\mathcal{Q}} \rangle_S$. 
This latter assumption yields

\[
  St Pr  \dfrac{\mathrm{d} \langle \xi' \rangle_S}{\mathrm{d} t} \sim
 Nu \dfrac{r_i r_o}{\rmeltmean^2},
\]
and allows us to derive the following scaling relation for the typical time of variation of the interface roughness
\begin{equation}
 \langle \tautopo \rangle_S \sim \dfrac{St Pr}{Nu} \rmeltmean^2 \langle \xi' 
\rangle_S\,.
\label{eq:tautopo_scaling}
\end{equation}
The proportionality between the correlation time and the Stefan number is expected since large $St$ implies large latent heat and therefore slower melting dynamics.
This is also consistent with previous observation of melting rates being inversely proportional to the Stefan number \citep{Favier19}.
Figure~\ref{fig:tau_topo} shows the validity of this scaling relation, with little scatter in the data and a slope close to the 
expected value of one.
This a posteriori validates the assumption  $\langle 
\mathcal{Q}' \rangle_S\langle /\tilde{\mathcal{Q}} \rangle_S \approx \text{const.}$ retained in the derivation.

\subsection{Roughness wavelength and amplitude}

\begin{figure}
 \centering
 \includegraphics[width=.49\textwidth]{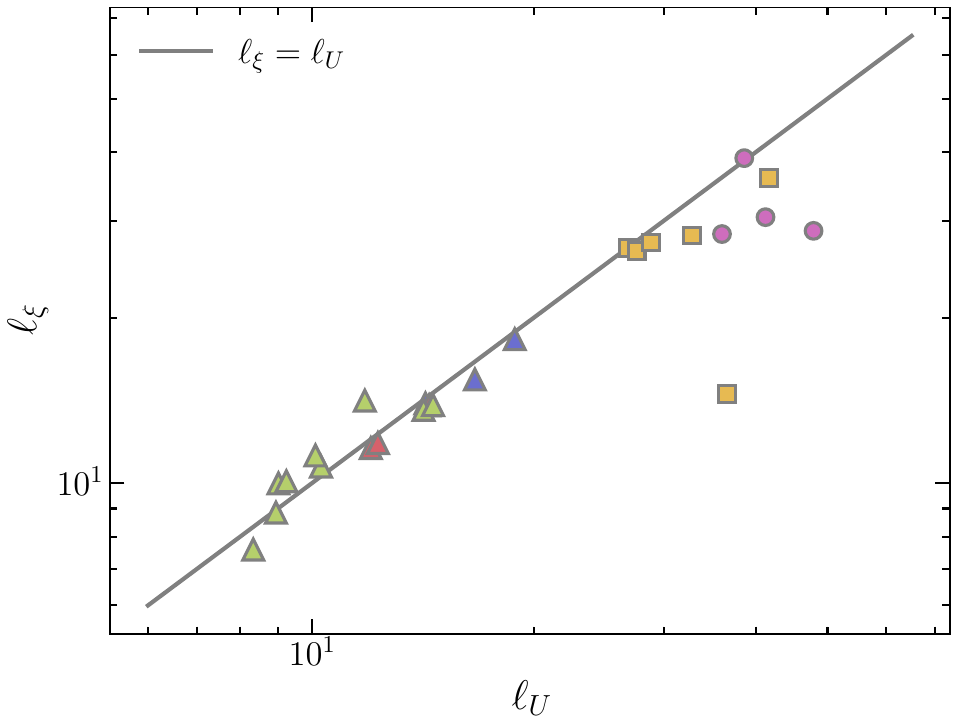}
 \caption{Comparison between the convective flow wavenumber
$\ell_U$ and the roughness wavenumber $\ell_\xi$.  The solid line corresponds to the equality $\ell_\xi=\ell_U$. Symbols carry the same meaning as in Fig.~\ref{fig:parameters}.}
\label{fig:ells}
\end{figure}

To evaluate the horizontal size of topography, Fig.~\ref{fig:ells} shows a comparison between the typical sizes of the convective flow and topography, both quantities being evaluated by the spherical harmonic degree where the corresponding spectra reach their maximum values (Eq.~\ref{eq:ellU} and 
Eq.~\ref{eq:ellXi}). 
The two wavenumbers are found to be broadly similar, indicating
that the horizontal scale of topography follows that of
the underlying convective pattern, in agreement with previous findings by e.g. \citet{Rabba18} or \citet{Favier19}.
There is less of an agreement at larger degrees (smaller lengthscales), for which several outliers exhibit topographic changes of larger lengthscale than that of the convective flow.
We recall that the Stefan number has been fixed to unity in all our simulations, whereas larger values 
could yield a smoothing of the small-scale topography variations.
Given the moderate changes of $\ell_U$ across the parameter space studied here, we can however merely speculate on the possible increasing disagreement between $\ell_\xi$ and $\ell_U$ for smaller-scale convective flows.

\begin{figure*}
 \centering
 \includegraphics[width=0.95\textwidth]{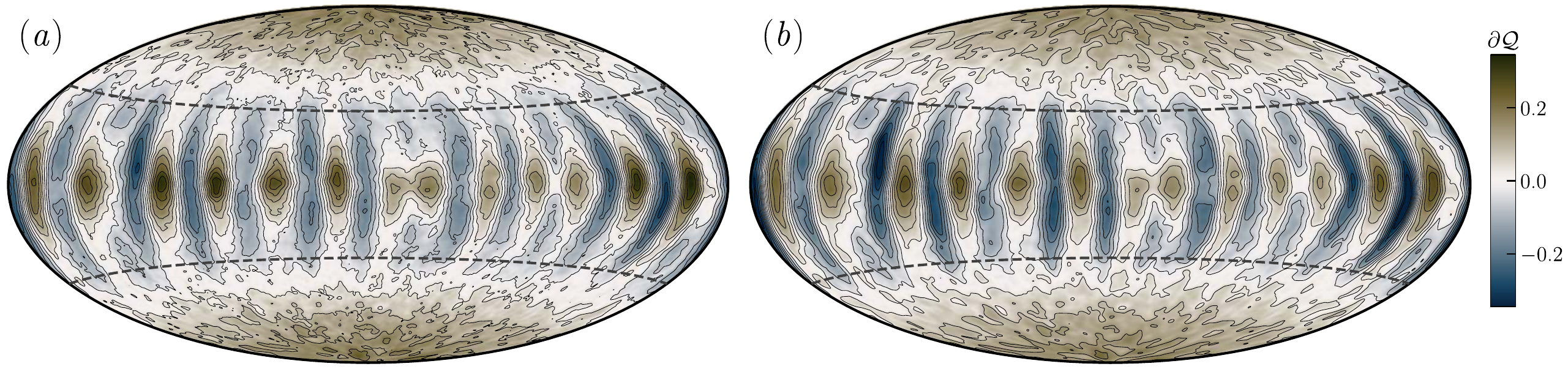}
 \caption{ Comparison between the actual time-averaged heat flux fluctuations 
$\partial \mathcal{Q}$ (panel \textit{a}) and the 
theoretical model expressed in Eq.~\eqref{eq:topo_theo} (panel \textit{b}) for 
a simulation with $E=10^{-3}$, $Ra=10^7$ and $\tmelt=0.6$. This is the same 
configuration as the one 
shown in Fig.~\ref{fig:rmelts}(\textit{b}) and Fig.~\ref{fig:rmelt_tau}.
In each panel, the dashed lines mark the location of the effective tangent 
cylinder.}
\label{fig:flux_theo}
\end{figure*}

%

For the configurations where the correlation time of topography exceeds the mean thermal diffusion time in the solid phase, i.e. $\tau_\xi > Pr\,\hS^2$, the roughness wavelength and amplitude can be related to the heat flux fluctuations using the perturbative model expressed in Eq.~\eqref{eq:topo_theo}. Figure~\ref{fig:flux_theo} shows a comparison between the actual time-averaged heat flux variations $\partial \mathcal{Q}$ (panel \textit{a}) and the model (panel \textit{b}) for the same numerical simulation already discussed in Fig.~\ref{fig:rmelt_tau}.
Most of the heat flux variations are localised outside the tangent cylinder and locked in the columnar topography (Fig.~\ref{fig:rmelts}\textit{b}).
The model --which we recall only retains the first-order contributions in terms of amplitude of topography-- manages to accurately reproduce the observed heat flux variations.
In this configuration, the  non-axisymmetric topography promotes heat flux heterogeneities that can reach up to $20\%$ of the average heat flux. 
At this stage, it is however unclear whether the locking phenomenon exemplified here will persist at stronger convective forcing \citep[see e.g.][]{Yang23} or is promoted by the large-scale convective pattern which develops at the moderate supercriticalities considered here.

Given the predominantly columnar nature of the interface roughness in this configuration, one can tentatively approximate the topographic changes by one single sectoral spherical harmonic function of degree and order $\ell_\xi$. Within this limit, the heat flux variations in the equatorial plane are approximated by
\begin{equation}
\begin{aligned}
 \max_{\vartheta=\pi/2,\varphi}|\partial \mathcal{Q}| & \approx 
   2 \dfrac{|\overline{\xi_{\ell_\xi \ell_\xi}}|}{\rmeltmean}
 f_{\ell_\xi}(\eta_S) 
 \max_{\vartheta=\pi/2,\varphi}|Y_{\ell_\xi \ell_\xi}|, \\
 & \approx 
   2 \dfrac{|\overline{\xi_{\ell_\xi \ell_\xi}}|}{\rmeltmean}
 \dfrac{f_{\ell_\xi}(\eta_S)}{2^{\ell_\xi} 
\ell_\xi!}\sqrt{\dfrac{(2\ell_\xi+1)!}{4\pi}}\,.
\label{eq:topo_theo_single_mode}
\end{aligned}
\end{equation}
For the model with $E=10^{-3}$, $Ra=10^7$ and $\tmelt=0.6$, one gets 
$\ell_\xi=11$ and $|\overline{\xi_{\ell_\xi  \ell_\xi}}|/\rmeltmean \approx 3.4 \times 
10^{-3}$ which yields  $\max_{\vartheta=\pi/2,\varphi}|\partial \mathcal{Q} |
\approx 0.1$, a value slightly underestimated compared to the actual extrema of the heat flux variations in the equatorial plane 
(Fig.~\ref{fig:flux_theo}\textit{a}). 
Despite the rather crude single mode 
approximation involved in Eq.~\eqref{eq:topo_theo_single_mode}, this equation allows to directly relate the amplitude and dominant wavelength of the interface topography to the corresponding heat flux heterogeneities,
which can prove useful for further order of magnitude estimates in the relevant 
geophysical regime.


\section{Geophysical implications}
\label{sec:geophy}

\begin{figure}
 \centering
  \includegraphics[width=0.49\textwidth]{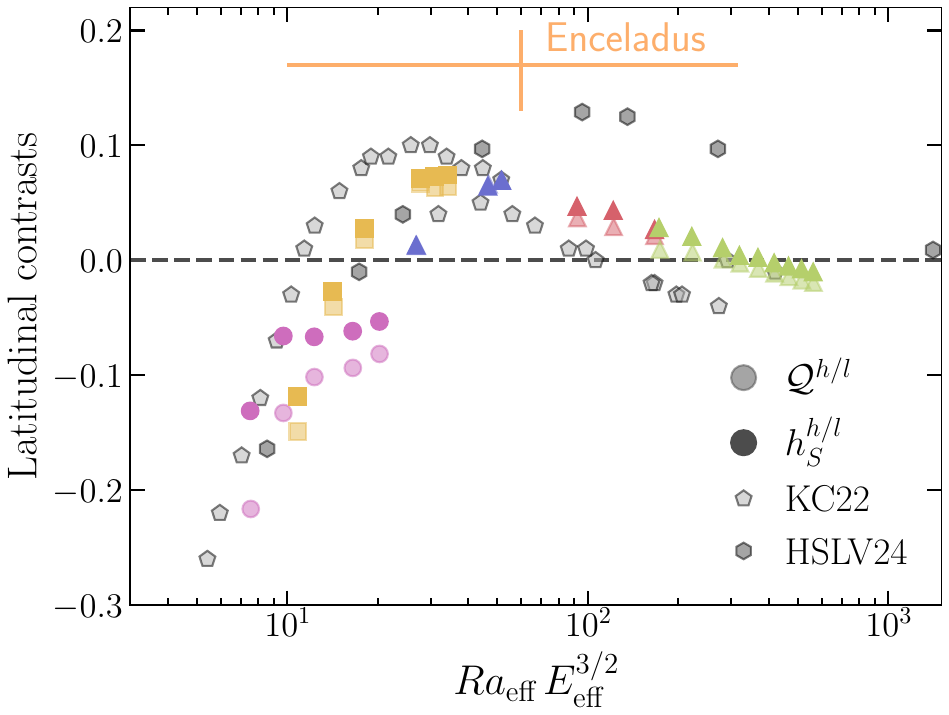}
  \caption{Relative contrasts of ice thickness ($\hshl$, 
Eq.~\ref{eq:hshl}) and heat fluxes ($\qhl$, Eq.~\ref{eq:qhl}) inside and 
outside the effective tangent cylinder as a function of $\RaL\,\EL^{3/2}$. The 
relative changes of 
the size of the solidus $\hshl$ (high-opacity symbols) have been measured 
in the simulations with a phase change while the heat flux 
contrasts $\qhl$ (low-opacity symbols) come from the equivalent 
purely convective models. The symbols carry the same meaning as in 
Fig.~\ref{fig:parameters}. For comparison purposes, the measures of $\qhl$ coming
from the monophasic simulations from 
\citet{Kvorka22} with rigid boundaries (here KC2022) and from 
\citet{Hartmann24} with $r_i/r_o=0.8$ (here HSLV24)  have been included. To 
account for the difference in the definition of the gravity profile, the 
Rayleigh numbers provided by \citet{Kvorka22} have been rescaled by the average 
of their gravity profile over 
ours. For Enceladus, estimates of $\hshl$ come from \citet{Cadek19}, while the 
uncertainties on $E$ and $Ra$ come from \citet{Soderlund19}.}
\label{fig:heatflux_rmelt}
\end{figure}

We already noticed earlier that the 
ice shell is either thicker in the equatorial or in the polar regions  
depending on the influence of rotation (Fig.~\ref{fig:theta}).
Earlier studies of monophasic rotating convection by e.g. 
\citet{Amit20} and \citet{Kvorka22} characterised two heat flux regimes termed 
``polar'' and ``equatorial cooling'' in which the heat flux is respectively 
larger in 
the polar or equatorial regions. The two studies however disagree on the 
control parameter which governs the transition between these two regimes. 
\citet{Amit20} advocate 
that polar cooling occurs when $Ra\,E^{8/5} > 10$, while \citet{Kvorka22} 
obtain this regime whenever $1<Ra\,E^{12/7} < 10$. This latter 
combination of exponents on $Ra$ and $E$ is obtained by assuming a transition 
between the heat transfer of rapidly-rotating convection $Nu\sim Ra\,E^{3/2}$ 
\citep[e.g.][]{Julien12} and its non-rotating RBC counterpart $Nu_\text{NR} \sim Ra^{1/3}$
\citep[see][]{Gastine16}.
The differences in the scaling exponents between the two studies could 
arise from the different adopted mechanical boundary conditions.

Given the different regionalized dynamics in spherical shell convection 
\citep{Wang21,Gastine23}, we find however more appropriate to define
this transition in terms of the scaling behaviour of polar convection.
In the limit of rapid rotation and moderate supercriticalities, the heat 
transfer in the polar regions of 
spherical shells with rigid boundaries closely follows the scaling behaviour
obtained in planar simulations, i.e. $Nu \sim 
Ra^3 E^4$ \citep{King12,Stellmach14,Gastine23}.
Assuming that the transition between equatorial and polar cooling is defined by the crossing between this scaling law and $Nu_\text{NR}$ then yields a parameter combination $Ra\,E^{3/2}$.
This is in line with the recent study by \citet{Hartmann24} who report an enhancement of the polar heat transport in spherical shell convection whenever $Ra\,E^{3/2}>1$ using a set of numerical simulations with a fixed Rayleigh number of $Ra=10^6$ and various Ekman numbers and radius ratios.

In terms of diagnostics, both regimes are usually characterised by defining the relative 
differences of the axisymmetric heat flux inside and outside the tangent cylinder 
\citep[e.g.][]{Amit20,Kvorka22,Bire22}:
\begin{equation}
 \qhl = 
\dfrac{[\tilde{Q}]_{0}^{\vartheta_\text{TC}}-
[\tilde{Q}]_{\vartheta_\text{TC}}^{\pi-\vartheta_\text{TC}}}{
[\tilde{Q}]_{0}^{\vartheta_\text{TC}}+
[\tilde{Q}]_{\vartheta_\text{TC}}^{\pi-\vartheta_\text{TC}}},
\label{eq:qhl}
\end{equation}
where $\vartheta_\text{TC}=\arcsin \etaL$ in the phase field models and 
$\vartheta_\text{TC}=\arcsin \eta$ without phase change.
In the above expression, the square brackets have been employed to define angular averages between two colatitudes
\[
 [f]_{\vartheta_1}^{\vartheta_2} =  \dfrac{1}{S_{\vartheta_1}^{\vartheta_2}}
\int_{\vartheta_1}^{\vartheta_2} f(\vartheta)\sin\vartheta 
\,\mathrm{d}\vartheta, \quad S_{\vartheta_1}^{\vartheta_2}= 
\int_{\vartheta_1}^{\vartheta_2}\sin\vartheta 
\,\mathrm{d}\vartheta\,.
\]
Here we accordingly define the relative differences in the axisymmetric ice 
thickness:
\begin{equation}
  \hshl=-\dfrac{[\tilde{\xi}]_{\vartheta_\text{TC}}^{\pi-\vartheta_\text{TC}}
-[\tilde{\xi}]_{0}^{\vartheta_\text{TC}}
}{2r_o-
[\tilde{\xi}]_{0}^{\vartheta_\text{TC}}
-[\tilde{\xi}]_{\vartheta_\text{TC}}^{\pi-\vartheta_\text{TC}}}\,,
\label{eq:hshl}
\end{equation}
where the minus sign has been introduced to ensure that the variations of 
$\hshl$ carry the same sign as those in $\qhl$.

Figure~\ref{fig:heatflux_rmelt} shows $\hshl$ for the simulations with a phase 
change (coloured high-opacity symbols) and $\qhl$ for the corresponding models 
without phase change (coloured low-opacity symbols) as a function of 
$\RaL\,\EL^{3/2}$. For comparison purposes, the simulations from 
\citet{Kvorka22} with rigid boundaries and from \citet{Hartmann24} with 
$r_i/r_o=0.8$ have been included. Both diagnostics 
follow similar trends. When the influence of rotation is the strongest, i.e. 
$\RaL\,\EL^{3/2} \lesssim 10$, the heat flux flux is larger outside the tangent 
cylinder and the ice is thinner there.
For a limited range of parameters, $10\lesssim \RaL\,\EL^{3/2} \lesssim 100$,
the heat flux is larger in the polar regions and ice is accordingly thicker 
outside the tangent cylinder. For larger values, both diagnostics
taper off as the influence of rotation diminishes and the flow gradually loses 
its preferred axis.
The limited number of simulations as well as the remaining scatter however 
prevent us to ascertain that $\RaL\,\EL^{3/2}$ is the right parameter 
combination to describe the transition. 
In addition, recent simulations by \citet{Song24} 
show that the heat transfer scaling rather adheres to $Nu\sim 
Ra^{3/2}E^2 Pr^{-1/2}$ in Cartesian geometry with rigid boundaries for $E < 
3\times 10^8$ and $Ra\,E^{4/3} > 10$. This is in line with the findings by 
\citet{Stellmach14}
who also report the same diffusivity-free scaling of rotating convection 
whenever stress-free boundary conditions are employed. In the geophysical 
regime relevant to the icy satellites, this is hence plausible that the 
transition between polar and equatorial cooling is rather governed by a 
$Ra\,E^{12/7}$ parameter combination.

With all these possible caveats in mind, we nevertheless tentatively locate in 
Fig.~\ref{fig:heatflux_rmelt} the expected values for Enceladus for which ice 
shell thickness models have been devised. Using topography and gravity data 
from Cassini, \citet{Cadek19} for instance derived a model of Enceladus' 
ice shell thickness which is about $30$~km thick at the equator and reaches 
$15$ ($5)$~km near the North (South) pole. This would place this subsurface 
ocean in the polar cooling regime with $\hshl\in[0.13,0.20]$.
Due to the uncertainties on the ocean thickness, the estimated values for $E$ 
and $Ra$ respectively span the intervals $[10^{-10},10^{-11}]$ and 
$[10^{16},10^{19}]$ \citep[e.g.][]{Soderlund19,Cabanes24}. 
These estimates yield $Ra\,E^{3/2}>10$ a parameter combination indeed 
compatible with the polar cooling regime obtained in numerical models. The 
amplitude of $\hshl$ for 
Enceladus however exceeds the largest contrast obtained in our simulations as 
well as --to a lesser extent-- the largest heat flux contrasts $\qhl$ obtained 
by \citet{Kvorka22} and \citet{Hartmann24} for their $r_i/r_o=0.8$ simulations.
Simulations by \citet{Bire22} conducted at lower Ekman numbers also suggest
a gradual drop  of $\qhl$ when the radius ratio of the ocean increases
(see their Fig.~12).
At this stage, this is hence unclear whether the intrinsic heat flux 
fluctuations of rotating convection are susceptible to reach a sufficient 
amplitude at the geophysical parameters to explain the topographic changes of 
Enceladus' ice. The strong North-South asymmetry could also originate from 
large scale basal heat flux heterogeneities or from temperature contrasts at 
Enceladus' surface
\citep[e.g.][]{Lemasquerier23}. Ice shell models of Titan are more uncertain but also
suggestive of a polar cooling regime, with for instance $\hshl \sim 0.05$ in the 
model by \citet{Lefevre14}. Due to the least rotational constraint on Titan's 
subsurface oceanic flows, estimates of $Ra$ and $E$ from \citet{Soderlund19} 
yield $Ra\,E^{3/2} \sim \mathcal{O}(10^4)$. Because of the scatter of the 
numerical simulations beyond $Ra\,E^{3/2} \sim 100$ 
(Fig.~\ref{fig:heatflux_rmelt}), it is uncertain whether such 
large scale topographic changes could be attributed to the convective 
fluctuations or are rather promoted by large scale thermal heterogeneities at 
the base or at the top of the fluid layer.
At this stage, it is also important to recall that several physical ingredients such as salinity and thermobaric
effects have been neglected in the current model and are likely to
change the interplay between heat flux changes and mean topography
\citep[e.g.][]{Kang23}.

Despite all the shortcomings of our models, the observation of long-lived stable non-axisymmetric corrugations of the solid-liquid interface in several simulations prompts us to attempt a 
re-scaling to the relevant geophysical regime. To do so, we make the following 
bold assumptions: (\textit{i}) the non-axisymmetric topography can be described 
by one single sectoral model of size $\ell_U$; (\textit{ii}) the ice layer is in 
a quasi-equilibrated conducting state that allows the application of the 
model Eq.~\eqref{eq:topo_theo} which relates the heat flux variations to the 
topographic wavelength and amplitude. Given the estimated average radius ratio 
of the ice layers of the icy satellites, the single mode approximation (Eq.~\ref{eq:topo_theo_single_mode}) can be further simplified to
\begin{equation}
 \partial \xi \sim \dfrac{\partial \mathcal{Q}}{\ell_U} \sim \dfrac{\partial 
\mathcal{Q} {\mathcal{L}_U}}{\pi \rmeltmean},
\end{equation}
where $\mathcal{L}_U$ is the dominant convective flow lengthscale.
The studies of the transfer functions of basal heterogeneous heat fluxes 
through a rotating convective layer carried out by \citet{Terra23} and 
\citet{Lemasquerier23} suggest
$\partial \mathcal{Q} \sim 2-4$ for Enceladus \citep[see also][for similar 
estimates]{Cadek19} and $\partial \mathcal{Q} \sim 0.1-0.5$ for Titan 
\citep{Choblet17}.
To provide an estimate of the lengthscale $\mathcal{L}_U$,
we further assume that the oceanic flows are rotationnally-constrained
and that the kinetic energy spectra of the zonal jets
and the non-axisymmetric residuals reach their maxima at similar scales,
in practice found by previous studies to be close to
the Rhines scale \citep[e.g][]{Guervilly19,Lemasquerier23a}.
We therefore
use the jet sizes derived by \citet{Cabanes24} to estimate $\mathcal{L}_U$, such
that $\mathcal{L}_U \sim 0.1D \sim 2$~km 
for Enceladus and $\mathcal{L}_U \sim 0.5D\sim 100$~km for Titan, with $D$ the 
ocean depth.
For both icy satellites, this yields $\partial\xi \sim 
\mathcal{O}(10^{-3}-10^{-2})$, depending on the amplitude of the heat flux 
variations. This translates to non-axisymmetric topographic changes at the
solid-liquid interface that could 
reach up to $\mathcal{O}(10^2-10^3)$~meters for Enceladus and
$\mathcal{O}(10^3-10^4)$~meters for Titan. 
Considering 2-D thermo-mechanical models of the ice response 
to basal heat flux variations, \citet{Kihoulou23} demonstrated that the topographic
changes in a conducting ice shell strongly depend on the ice viscosity as well as on
the variations of the melting temperature with pressure (see their Fig.~6).
Despite these uncertainties, their estimates of topographic changes are lower than ours,
typically covering the range of $\mathcal{O}(10^2-10^3)$~m
at the ice-water interface \citep[see also][]{Kang23}.
Again, accounting for double-diffusive effects, ice creeping
and temperature variations due to pressure changes along the deformed boundary
are susceptible to change these conclusions.

\section{Conclusion}
\label{sec:conclu}

Improving understanding of the dynamical coupling between the ocean and the overlying
ice layer is of primary importance to better characterise the hydrosphere of
the icy moons. In this study we have developed a novel approach to model
rotating convection with a phase change using a phase field formulation \citep[e.g.][]{Beckermann99}. 

To examine the interplay between rotating convection and a melting boundary,
we have conducted a series of numerical simulations in spherical
geometry varying the control parameters --the Rayleigh number $Ra$ and 
the Ekman number $E$-- as well as the melting temperature.
We have split the analysis between the large-scale axisymmetric topography
and the non-axisymmetric features. 
For the former and in line with previous monophasic convection models, we have evidenced two
regimes in which the mean axisymmetric heat flux either peaks at the equator 
or at the poles with a corresponding regionalized ice thinning. Transition between
the two happens when $Ra\,E^{3/2} \approx 10$, a parameter combination which
is obtained when the scaling behaviour of rotating convection in the polar
regions of the spherical shell crosses the one of classical RBC \citep{Hartmann24}.

We have conducted numerical simulations  of the equivalent monophasic 
setups and we have derived a perturbative model of the heat equilibrium in a quasi-spherical 
shell with bottom topographic changes. 
This enabled us to show that the relative heat flux variations 
at the top of the monophasic simulations 
usually provide a good guess of the actual flux as long as the mean axisymmetric
topographic changes are small.
As expected, departures between the configuration with a phase
change and their counterparts without become more pronounced for large topographic variations.

Non-axisymmetric topography, termed roughness in this study, yields significant differences
with the usual rotating convective flow in spherical shells. In particular, we have evidenced
the formation of long-lived large-scale columnar troughs and crests which develop
for intermediate rotational constraints.
Convective upwellings are then locked in the topographic changes of the solid-liquid interface. 
Although 
our parameter coverage does not allow to determine the parameter combination which governs the transition to locked-in convection \citep[see also][]{Yang23}, a tentative rescaling of the amplitude
of this non-axisymmetric topography to the planetary regime 
yields $\mathcal{O}(10^2-10^3)$ meters for Enceladus and $\mathcal{O}(10^3-10^4)$ meters for Titan.
Those values are substantially larger than current estimates by e.g.~\citet{Corlies17} or \citet{Kang23}.
These differences could possibly arise from several simplifications of our model.
We do not include the effect of salinity \citep{Ashkenazy21,Wong22}, the
pressure dependence of the melting temperature \citep{Labrosse18,Lawrence24}, 
the creep properties of ice
\citep[e.g.][]{Weertman83,Shibley24}
the effect of tidal heating in the ice crust \citep{Beuthe19,Behounkova21},
or the thermal heterogeneities at the base of 
the ocean \citep{Terra23,Lemasquerier23} or at the moon's 
surface \citep{Weller19}. All those endogenic physical processes 
are possible candidates susceptible to raise the complexity
of the results described here. Accounting for those effects could be the subject 
of future studies using the phase change formalism discussed here.


On a longer term, improving our understanding of the interplay between oceanic flows and phase changes
in the icy moons will also require to conduct numerical simulations at more extreme parameters. 
Though numerically challenging 
\citep[e.g.][]{Song24}, this would allow to consolidate the scaling relation which governs the 
transition between equatorial and polar cooling and to study the coupling between mean zonal
flows and topography \citep[e.g.][]{Hay23}.



\section*{Acknowledgements}

The authors thank two anonymous reviewers for their suggestions that helped improve the manuscript.
The authors acknowledge the support of the French Agence 
Nationale de la Recherche (ANR), under grant ANR-20-CE49-0010 (project COLOSSe).
Numerical computations have been carried out on the \texttt{S-CAPAD/DANTE} platform at IPGP.
All the figures have been generated using \texttt{matplotlib} \citep{Hunter07}
and \texttt{paraview} (\url{https://www.paraview.org}). Several colormaps come from
the \texttt{cmocean} package by \cite{cmocean}.

\appendix

\section{Mapping functions for Chebyshev collocation method}

\label{sec:mappings}

When the fluid domain features regions of rapid changes, considering mappings 
can significantly improve the convergence of the Chebyshev pseudo-spectral
approximation.
The mapping introduced by \cite{Kosloff93} aims at reducing the grid points clustering near the boundaries inherent in the native Gauss-Lobatto collocation grid points.
The amplitude of the grid stretching is governed by a control parameter $\alpha_1\in[0,1[$.
The mapping function is expressed by
\begin{equation}
 \mathcal{F}(x)=\dfrac{\arcsin (\alpha_1 x)}{\arcsin \alpha_1}\,.
\end{equation}

The main purpose of the mapping defined by \cite{Bayliss92} is to refine the grid spacing around a particular interior point $x=\alpha_2$.
It is governed by two input parameters: $\alpha_1$ which controls the stiffness of the grid refinement and $\alpha_2\in[-1,1]$ which defines the center of the mapping function.
It is defined by
\begin{equation}
 \mathcal{F}(x)=\alpha_2+\dfrac{1}{\alpha_1}\tan[\lambda(x-x_0)],
 \label{eq:BT}
\end{equation}
with 
\[
  \lambda=\dfrac{\arctan[\alpha_1(1-\alpha_2)]}{1-x_0},\
x_0=\dfrac{\arctan[\alpha_1(1+\alpha_2)]-1}{\arctan[\alpha_1(1-\alpha_2)]+1}\,.
\]

Similarly to the mapping by \cite{Bayliss92}, the mapping by \cite{Jafari15} was also introduced to handle steep localized fronts.
It is governed by three input parameters: $\alpha_1$ and $\alpha_2$ retaining the same meaning as for the previous mapping and $\alpha_3 \in[0.2,0.9]$ being a small parameter that we keep to a fixed value of $\alpha_3=0.4$.
It is defined by
\begin{equation}
\mathcal{F}(x)=\alpha_2+\dfrac{1}{\alpha_1}\sinh\left\lbrace A\left[\dfrac{
\tan(x\arctan C)}C-1\right]+B \right\rbrace
\end{equation}
with
\[
\begin{aligned}
 A& =\dfrac{1}{2}\left\lbrace
 \argsinh[\alpha_1(1-\alpha_2)]+
 \argsinh[\alpha_1(1+\alpha_2)]
 \right\rbrace\,, \\
 B & = \argsinh[\alpha_1(1-\alpha_2)], \\
 C & 
=\left[\left|\Im\left\lbrace\dfrac{1}{A}\left(\dfrac{\mathrm{i}\pi}{2}-B\right)+
1
\right\rbrace\right|+\alpha_3\right]^{-1}\,.
\end{aligned}
\]


\section{A benchmark for rotating convection in a spherical shell with a phase change}
\label{sec:bench}

To validate the numerical implementation of the phase field method in the pseudo-spectral code \texttt{MagIC}, we consider a weakly nonlinear configuration of rotating convection close to onset. Similarly to classical benchmarks of monophasic convection in spherical
geometry \citep[e.g.][]{Christensen01,Marti14}, this allows to reach a saturated
state which takes the form of steadily drifting convective columns. This quasi-stationarity allows to benchmark well-defined integrated diagnostics such as the total kinetic energy $E_K$ or the Nusselt number $Nu$ defined by \eqref{eq:ek} and \eqref{eq:nu} respectively.
Here we adopt $E=10^{-3}$, 
$Ra=1.8\times 10^5$, $Pr=1$, $St=1$, $\tmelt=0.25$ for a radius ratio $\eta=0.35$. For this parameter combination, the solid-liquid 
interface equilibrates around the mean radius $\rmeltmean \approx 1.09$, which yields the following 
effective quantities: $\EL\approx 3.4\times 10^{-3}$, $\RaL\approx 1.5\times 10^4$ and $\etaL\approx 
0.5$. For this Ekman number and radius ratio, the first unstable mode that becomes linearly unstable 
in the equivalent monophasic convection problem features an azimuthal wavenumber $m=5$ and a critical
Rayleigh number $Ra_c =1.322\times 10^4$ \citep{Barik23}. This is closely followed by the $m=6$ mode 
which has $Ra_c=1.341\times 10^4$. In order to define a reproducible reference case, we hence 
initiate all the numerical models using a thermal perturbation with a fivefold azimuthal symmetry. 

\begin{figure*}
 \centering
 \includegraphics[width=0.95\textwidth]{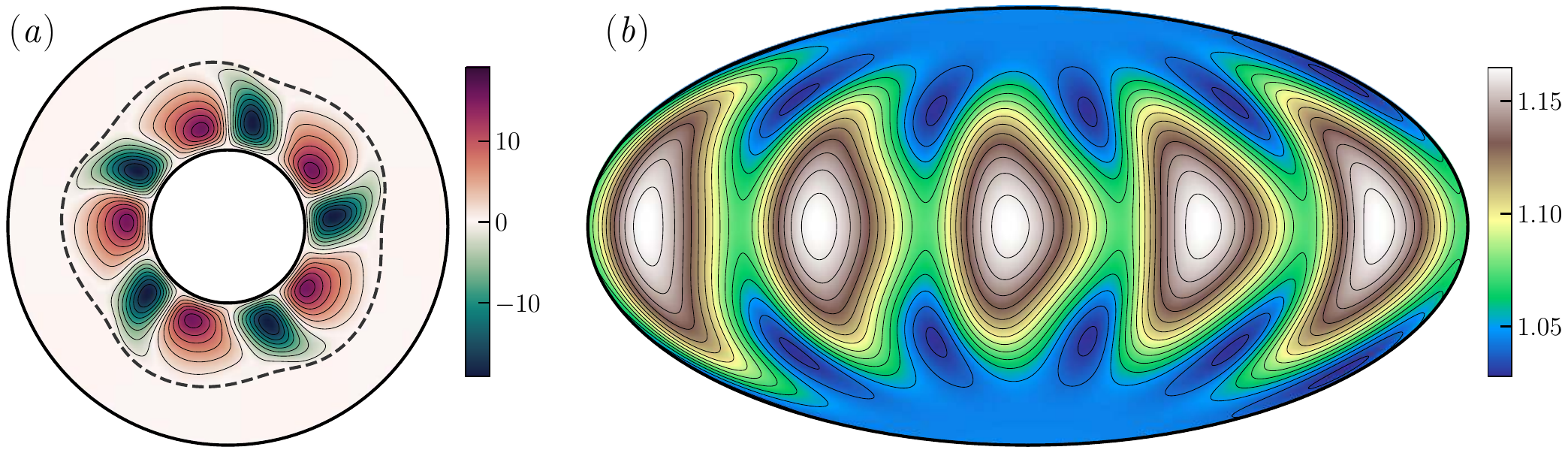}
 \caption{Snapshots of the benchmark configuration with $E=10^{-3}$, $Ra=1.8\times 10^5$, $Pr=1$, $St=1$, $\tmelt=0.25$ and $r_i/r_o=0.35$ for a numerical simulation with $\epsilon = 10^{-3}$. (\textit{a}) Radial velocity $u_r$ in the equatorial plane. (\textit{b}) Hammer projection of the melt radius $\rmelt$. The dashed line in panel (\textit{a}) marks the location of the solid-liquid interface $\rmelt(\vartheta=\pi/2,\varphi)$.}
 \label{fig:bench}
 \includegraphics[width=0.95\textwidth]{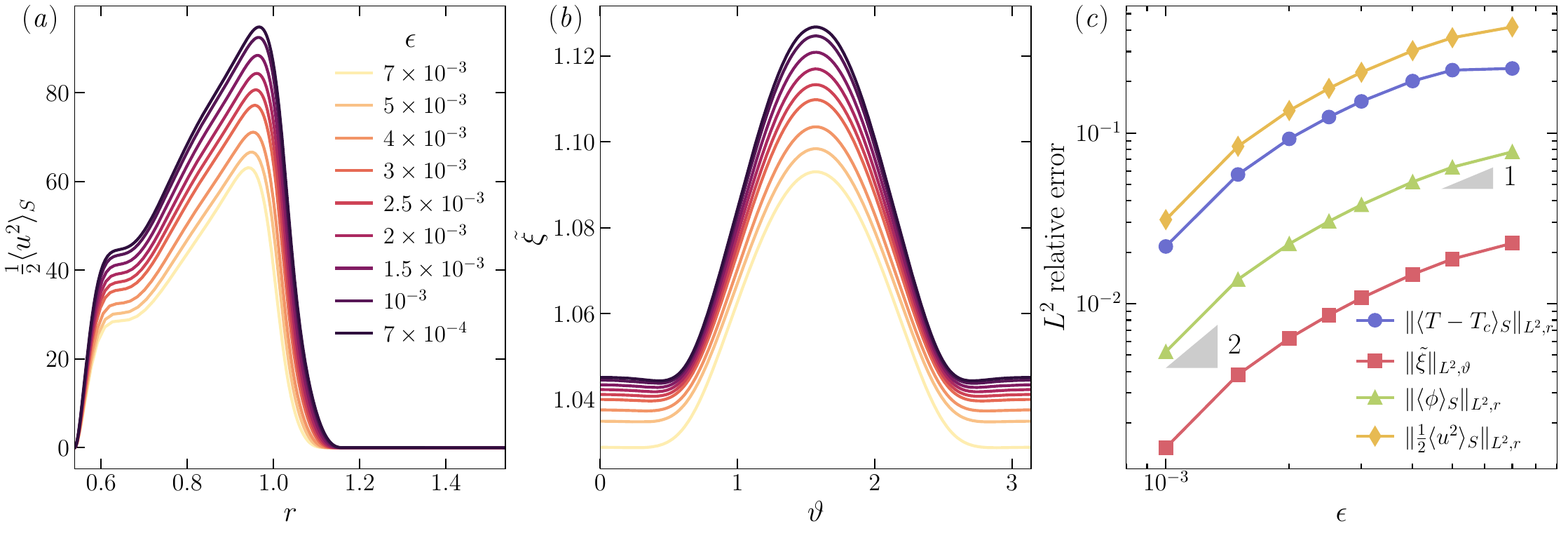}
 \caption{(\textit{a}) Comparison of the radial profiles of kinetic energy for the benchmark configuration with decreasing
 values of the Cahn number $\epsilon$. (\textit{b}) Comparison of the azimuthal average of the
 melt radius $\tilde{\xi}$ for decreasing values of $\epsilon$. (\textit{c})  $L^2$ relative errors
 (Eq.~\ref{eq:L2_errors}) of several diagnostic quantities as a function of $\epsilon$.}
\label{fig:convergence_bench}
\end{figure*}

Figure~\ref{fig:bench} shows a snapshot of the radial velocity (panel \textit{a}) and the 
solid-liquid interface (panel \textit{b}) for a numerical simulation with $\epsilon = 10^{-3}$.
The solution takes the form of convective columns confined in the lower half of the fluid
domain. The interface features a fivefold symmetry corrugation outside the effective tangent cylinder.
Due to the localised convective columns, the ice thickness is thinner near the equator
than the poles. 

Figure~\ref{fig:convergence_bench} illustrates the convergence of the phase 
field method for this benchmark case when decreasing the Cahn number $\epsilon$.
We recall that the phase-field formulation converges towards the original Stefan problem as $\epsilon$ vanishes.
Panel (\textit{a}) 
shows radial profiles of the kinetic energy, while panel (\textit{b}) shows the latitudinal profile 
of the axisymmetric melt radius $\tilde{\xi}$. On decreasing $\epsilon$, the solid-liquid transition
steepens, which goes along with a gradual increase of the kinetic energy content and an outward shift
of the interface. Both diagnostics are suggestive of a gradual convergence towards an asymptotic solution as $\epsilon \rightarrow 0$.
To examine this convergence in a more quantitative way, we compute in
Fig.~\ref{fig:convergence_bench}(\textit{c}) the $L^2$ relative error between a given phase field model and the reference solution, here defined by the simulation with the lowest value
of $\epsilon=7\times 10^{-4}$. We consider several diagnostics, either based on mean radial profiles
or mean latitudinal profiles:
\begin{equation}
\begin{aligned}
\|f\|_{L^2,r}& =\left(\dfrac{\int_{r_i}^{r_o}\left[f(r)-f_\text{ref}(r)\right]^2r^2\mathrm{d}r}{%
\int_{r_i}^{r_o}f_\text{ref}^2\mathrm{d}r}\right)^{1/2}\,, \\
\|f\|_{L^2,\vartheta}&=\left(\dfrac{\int_{0}^{\pi}\left[f(\vartheta)-f_\text{ref}(\vartheta)\right]^2 \sin\vartheta\mathrm{d}\vartheta}{%
\int_{0}^{\pi}f_\text{ref}^2\sin\vartheta\mathrm{d}\vartheta}\right)^{1/2}\,,
\end{aligned}
\label{eq:L2_errors}
\end{equation}
where $f_\text{ref}$ in the above equations correspond to the reference solution.
All diagnostics exhibit a similar trend with a slow convergence 
close to first order for the largest values of $\epsilon$ and a gradual increase
towards the second order for $\epsilon \approx 10^{-3}$.
We note that a similar transition from first to second order convergence has been observed in 
\cite{Favier19}. The convergence of our phase field model is therefore more complex than the second 
order reported by \citet{Hester20}.
In addition to the known influence of the penalty coefficient $\tau_p$ on the convergence rate \citep{Hester21a}, it is very likely that
the influence of rotation, which was not considered in \citet{Hester20}, modifies the convergence
behaviour of the phase field formulation.

\begin{table}
\caption{Table of results of the benchmark configuration. All the simulations 
have been conducted using the BT mapping with $\alpha_1=10$, the third order 
SBDF3 time scheme and a penalty coefficient $\tau_p=0.6$.}
\centering
{\footnotesize 
\begin{tabular}{rrrrrr}
\toprule
$\epsilon$ & $\rmeltmean$ & $E_K$ & $Nu$ & $\delta t$ & $(N_R, \ell_\text{max})$ \\
\midrule
$7\times 10^{-3}$ & $1.063$ & $11.93$ & $1.112$ & $8\times 10^{-6}$ & $(161,213)$  \\
$5\times 10^{-3}$ & $1.068$ & $13.04$ & $1.111$ & $4\times 10^{-6}$ & $(161,213)$  \\
$4\times 10^{-3}$ & $1.072$ & $14.25$ & $1.116$ & $10^{-6}$ & $(193,213)$  \\
$3\times 10^{-3}$ & $1.076$ & $15.86$ & $1.123$ & $8\times 10^{-7}$ & $(257,213)$  \\
$2.5\times 10^{-3}$ & $1.078$ & $16.81$ & $1.127$ & $4\times 10^{-7}$ & $(257,213)$  \\
$2\times 10^{-3}$ & $1.081$ & $17.82$ & $1.132$ & $3\times 10^{-7}$ & $(257,213)$  \\
$1.5\times 10^{-3}$ & $1.083$ & $18.94$ & $1.138$ & $2\times 10^{-7}$ & $(321,320)$  \\
$10^{-3}$ & $1.085$ & $20.09$ & $1.144$ & $10^{-7}$ & $(385,426)$  \\
$7\times 10^{-4}$ & $1.087$ & $20.77$ & $1.147$ & $10^{-7}$ & $(513,533)$  \\
\bottomrule
\end{tabular}
}
\label{tab:bench_results}
\end{table}

To ease the future comparison of phase field models in spherical geometry, Table~\ref{tab:bench_results} lists the values of several integrated 
diagnostics for the benchmark configuration ($\rmeltmean$, $E_K$ and $Nu$) 
along with the temporal and spatial resolution employed to ensure an
appropriate convergence of the models.

\section{Table of results}

Table~\ref{tab:results} lists the input dimensionless numbers, the main global diagnostics as well as the numerical parameters of all the direct numerical simulations considered in this study.
All simulations have been computed using $\eta=r_i/r_o=0.8$, $Pr=1$ and $St=1$.

\begin{table*}[width=.8\textwidth]
\caption{Table of results. All the simulations have been computed with a phase field parameter $a=1$.}
\centering
{\footnotesize
\begin{tabular}{rrrrrrrrrrrr}
\toprule
$\tmelt$ & $\rmeltmean$ & $\EL$ & $\RaL$ & $Re_\mathcal{L}$ & $\NuL$ & $\ell_U$ & $\ell_\xi$ & $\epsilon$ & $\tau_p$ &$(N_R, \ell_\text{max})$  & $({\rm Map},\alpha_1,\alpha_2)$ \\
\midrule
\multicolumn{12}{c}{$E=10^{-3},\ Ra=3\times 10^{6}$} \\
$0.00$ & -- & $10^{-3}$ & $3\times 10^{6}$ & $297.3$ & $9.54$ & $12$ & -- & -- & -- & $(65,426)$ & ({\rm GL},--,--) \\
$0.44$ & $4.87$ & $1.325\times 10^{-3}$ & $1.073\times 10^{6}$ & $170.9$ & $6.66$ & $14$ & $13$ & $2\times 10^{-3}$ & $0.40$ & $(193,426)$ & $({\rm JVH},20,0.73)$ \\
-- & -- & $1.324\times 10^{-3}$ & $1.072\times 10^{6}$ & $174.8$ & $6.92$ &$14$ & -- & -- & -- & $(65,426)$ & ({\rm GL},--,--) \\
$0.49$ & $4.83$ & $1.455\times 10^{-3}$ & $8.416\times 10^{5}$ & $147.0$ & $5.94$ & $16$ & $15$ & $3\times 10^{-3}$ & $0.20$ & $(193,341)$ & $({\rm JVH},5,0.64)$ \\
-- & -- & $1.455\times 10^{-3}$ & $8.416\times 10^{5}$ & $154.6$ & $6.44$ &$16$ & -- & -- & -- & $(65,341)$ & ({\rm GL},--,--) \\
$0.69$ & $4.58$ & $2.972\times 10^{-3}$ & $1.663\times 10^{5}$ & $66.9$ & $3.88$ & $18$ & $18$ & $3\times 10^{-3}$ & $0.40$ & $(257,213)$ & $({\rm KTE},0.99,-)$ \\
-- & -- & $2.972\times 10^{-3}$ & $1.663\times 10^{5}$ & $68.4$ & $4.03$ &$21$ & -- & -- & -- & $(65,213)$ & ({\rm GL},--,--) \\
\midrule
\multicolumn{12}{c}{$E=10^{-3},\ Ra=10^{7}$} \\
$0.00$ & -- & $10^{-3}$ & $10^{7}$ & $559.9$ & $14.39$ & $9$ & -- & -- & -- & $(97,426)$ & ({\rm GL},--,--) \\
$0.46$ & $4.90$ & $1.241\times 10^{-3}$ & $3.825\times 10^{6}$ & $324.0$ & $9.48$ & $12$ & $11$ & $5\times 10^{-3}$ & $0.20$ & $(97,213)$ & $({\rm JVH},20,0.79)$ \\
-- & -- & $1.241\times 10^{-3}$ & $3.810\times 10^{6}$ & $350.9$ & $10.80$ &$12$ & -- & -- & -- & $(65,341)$ & ({\rm GL},--,--) \\
$0.60$ & $4.82$ & $1.502\times 10^{-3}$ & $2.094\times 10^{6}$ & $254.5$ & $8.48$ & $12$ & $11$ & $3\times 10^{-3}$ & $0.40$ & $(129,394)$ & $({\rm JVH},10,0.62)$ \\
-- & -- & $1.506\times 10^{-3}$ & $2.084\times 10^{6}$ & $263.2$ & $9.14$ &$13$ & -- & -- & -- & $(65,256)$ & ({\rm GL},--,--) \\
$0.69$ & $4.71$ & $1.993\times 10^{-3}$ & $1.037\times 10^{6}$ & $181.0$ & $6.90$ & $14$ & $13$ & $3\times 10^{-3}$ & $0.40$ & $(193,341)$ & $({\rm JVH},7,0.43)$ \\
-- & -- & $1.993\times 10^{-3}$ & $1.037\times 10^{6}$ & $187.9$ & $7.44$ &$13$ & -- & -- & -- & $(65,256)$ & ({\rm GL},--,--) \\
\midrule
\multicolumn{12}{c}{$E=10^{-3},\ Ra=3\times 10^{7}$} \\
$0.00$ & -- & $10^{-3}$ & $3\times 10^{7}$ & $998.4$ & $20.59$ & $7$ & -- & -- & -- & $(97,512)$ & ({\rm GL},--,--) \\
$0.40$ & $4.95$ & $1.117\times 10^{-3}$ & $1.508\times 10^{7}$ & $673.9$ & $14.96$ & $8$ & $7$ & $3\times 10^{-3}$ & $0.25$ & $(257,341)$ & $({\rm JVH},20,0.894)$ \\
-- & -- & $1.117\times 10^{-3}$ & $1.508\times 10^{7}$ & $725.9$ & $17.46$ &$8$ & -- & -- & -- & $(65,341)$ & ({\rm GL},--,--) \\
$0.45$ & $4.93$ & $1.150\times 10^{-3}$ & $1.321\times 10^{7}$ & $630.0$ & $14.34$ & $8$ & $8$ & $3\times 10^{-3}$ & $0.25$ & $(257,341)$ & $({\rm JVH},10,0.79)$ \\
-- & -- & $1.150\times 10^{-3}$ & $1.321\times 10^{7}$ & $678.6$ & $16.70$ &$8$ & -- & -- & -- & $(65,341)$ & ({\rm GL},--,--) \\
$0.50$ & $4.92$ & $1.194\times 10^{-3}$ & $1.130\times 10^{7}$ & $584.4$ & $13.64$ & $9$ & $9$ & $3\times 10^{-3}$ & $0.25$ & $(257,512)$ & $({\rm JVH},20,0.78)$ \\
-- & -- & $1.195\times 10^{-3}$ & $1.129\times 10^{7}$ & $626.9$ & $15.84$ &$8$ & -- & -- & -- & $(65,341)$ & ({\rm GL},--,--) \\
$0.55$ & $4.89$ & $1.255\times 10^{-3}$ & $9.400\times 10^{6}$ & $533.6$ & $12.88$ & $9$ & $10$ & $3\times 10^{-3}$ & $0.25$ & $(257,341)$ & $({\rm JVH},10,0.7)$ \\
-- & -- & $1.255\times 10^{-3}$ & $9.399\times 10^{6}$ & $571.9$ & $14.89$ &$8$ & -- & -- & -- & $(65,341)$ & ({\rm GL},--,--) \\
$0.60$ & $4.86$ & $1.338\times 10^{-3}$ & $7.543\times 10^{6}$ & $478.1$ & $12.04$ & $10$ & $10$ & $3\times 10^{-3}$ & $0.25$ & $(257,341)$ & $({\rm JVH},10,0.62)$ \\
-- & -- & $1.338\times 10^{-3}$ & $7.543\times 10^{6}$ & $513.2$ & $13.88$ &$9$ & -- & -- & -- & $(65,341)$ & ({\rm GL},--,--) \\
$0.65$ & $4.83$ & $1.469\times 10^{-3}$ & $5.694\times 10^{6}$ & $417.1$ & $11.06$ & $10$ & $11$ & $3\times 10^{-3}$ & $0.25$ & $(257,341)$ & $({\rm JVH},10,0.55)$ \\
-- & -- & $1.469\times 10^{-3}$ & $5.694\times 10^{6}$ & $447.3$ & $12.73$ &$10$ & -- & -- & -- & $(65,341)$ & ({\rm GL},--,--) \\
$0.69$ & $4.79$ & $1.604\times 10^{-3}$ & $4.384\times 10^{6}$ & $371.2$ & $10.63$ & $11$ & $14$ & $3\times 10^{-3}$ & $0.40$ & $(193,426)$ & $({\rm JVH},5,0.46)$ \\
-- & -- & $1.604\times 10^{-3}$ & $4.384\times 10^{6}$ & $391.0$ & $11.54$ &$11$ & -- & -- & -- & $(65,512)$ & ({\rm GL},--,--) \\
$0.75$ & $4.70$ & $2.056\times 10^{-3}$ & $2.391\times 10^{6}$ & $277.2$ & $8.81$ & $14$ & $13$ & $3\times 10^{-3}$ & $0.30$ & $(193,341)$ & $({\rm JVH},5,0.23)$ \\
-- & -- & $2.056\times 10^{-3}$ & $2.391\times 10^{6}$ & $294.1$ & $9.82$ &$12$ & -- & -- & -- & $(65,341)$ & ({\rm GL},--,--) \\
$0.80$ & $4.57$ & $3.041\times 10^{-3}$ & $1.035\times 10^{6}$ & $187.3$ & $6.86$ & $14$ & $13$ & $3\times 10^{-3}$ & $0.25$ & $(257,341)$ & $({\rm JVH},5,0.04)$ \\
-- & -- & $3.041\times 10^{-3}$ & $1.035\times 10^{6}$ & $197.8$ & $7.76$ &$13$ & -- & -- & -- & $(65,341)$ & ({\rm GL},--,--) \\
\midrule
\multicolumn{12}{c}{$E=3\times 10^{-4},\ Ra=1.2\times 10^{7}$} \\
$0.00$ & -- & $3\times 10^{-4}$ & $1.2\times 10^{7}$ & $517.6$ & $13.39$ & $20$ & -- & -- & -- & $(97,426)$ & ({\rm GL},--,--) \\
$0.44$ & $4.90$ & $3.681\times 10^{-4}$ & $4.848\times 10^{6}$ & $308.5$ & $9.52$ & $26$ & $26$ & $10^{-3}$ & $1.51$ & $(193,512)$ & $({\rm JVH},100,0.837)$ \\
-- & -- & $3.681\times 10^{-4}$ & $4.848\times 10^{6}$ & $322.2$ & $10.08$ &$25$ & -- & -- & -- & $(65,293)$ & ({\rm GL},--,--) \\
$0.49$ & $4.88$ & $3.908\times 10^{-4}$ & $4.015\times 10^{6}$ & $275.2$ & $8.78$ & $27$ & $26$ & $3\times 10^{-3}$ & $0.30$ & $(193,341)$ & $({\rm JVH},50,0.796)$ \\
-- & -- & $3.908\times 10^{-4}$ & $4.015\times 10^{6}$ & $292.8$ & $9.50$ &$26$ & -- & -- & -- & $(65,293)$ & ({\rm GL},--,--) \\
$0.54$ & $4.85$ & $4.187\times 10^{-4}$ & $3.245\times 10^{6}$ & $246.3$ & $8.25$ & $28$ & $27$ & $3\times 10^{-3}$ & $0.30$ & $(129,341)$ & $({\rm JVH},50,0.72)$ \\
-- & -- & $4.181\times 10^{-4}$ & $3.243\times 10^{6}$ & $259.9$ & $8.82$ &$28$ & -- & -- & -- & $(97,341)$ & ({\rm GL},--,--) \\
$0.69$ & $4.69$ & $6.244\times 10^{-4}$ & $1.163\times 10^{6}$ & $150.6$ & $6.34$ & $32$ & $28$ & $2\times 10^{-3}$ & $0.60$ & $(257,426)$ & $({\rm BT},10,0.41)$ \\
-- & -- & $6.244\times 10^{-4}$ & $1.163\times 10^{6}$ & $154.4$ & $6.41$ &$32$ & -- & -- & -- & $(65,341)$ & ({\rm GL},--,--) \\
$0.75$ & $4.56$ & $9.602\times 10^{-4}$ & $4.777\times 10^{5}$ & $97.3$ & $4.79$ & $36$ & $14$ & $2\times 10^{-3}$ & $0.60$ & $(257,426)$ & $({\rm BT},10,0.135)$ \\
-- & -- & $9.602\times 10^{-4}$ & $4.777\times 10^{5}$ & $98.3$ & $4.83$ &$38$ & -- & -- & -- & $(65,341)$ & ({\rm GL},--,--) \\
$0.80$ & $4.35$ & $2.511\times 10^{-3}$ & $8.613\times 10^{4}$ & $45.0$ & $2.71$ & $41$ & $35$ & $2\times 10^{-3}$ & $0.30$ & $(385,213)$ & $({\rm BT},5,-0.36)$ \\
-- & -- & $2.511\times 10^{-3}$ & $8.613\times 10^{4}$ & $42.4$ & $2.94$ &$42$ & -- & -- & -- & $(65,341)$ & ({\rm GL},--,--) \\
\midrule
\multicolumn{12}{c}{$E=3\times 10^{-5},\ Ra=2.5\times 10^{8}$} \\
$0.00$ & -- & $3\times 10^{-5}$ & $2.5\times 10^{8}$ & $1968.3$ & $28.87$ & $28$ & -- & -- & -- & $(129,512)$ & ({\rm GL},--,--) \\
$0.50$ & $4.93$ & $3.432\times 10^{-5}$ & $1.008\times 10^{8}$ & $1123.3$ & $19.85$ & $35$ & $28$ & $5\times 10^{-4}$ & $1.20$ & $(385,682)$ & $({\rm JVH},15,0.82)$ \\
-- & -- & $3.432\times 10^{-5}$ & $1.008\times 10^{8}$ & $1143.0$ & $18.98$ &$33$ & -- & -- & -- & $(129,512)$ & ({\rm GL},--,--) \\
$0.59$ & $4.90$ & $3.674\times 10^{-5}$ & $7.416\times 10^{7}$ & $905.4$ & $17.42$ & $38$ & $38$ & $10^{-3}$ & $0.60$ & $(257,512)$ & $({\rm JVH},20,0.79)$ \\
-- & -- & $3.674\times 10^{-5}$ & $7.416\times 10^{7}$ & $952.0$ & $16.86$ &$35$ & -- & -- & -- & $(97,512)$ & ({\rm GL},--,--) \\
$0.69$ & $4.83$ & $4.344\times 10^{-5}$ & $4.297\times 10^{7}$ & $669.8$ & $13.98$ & $41$ & $30$ & $10^{-3}$ & $0.40$ & $(257,512)$ & $({\rm JVH},20,0.633)$ \\
-- & -- & $4.347\times 10^{-5}$ & $4.294\times 10^{7}$ & $702.3$ & $13.87$ &$40$ & -- & -- & -- & $(97,512)$ & ({\rm GL},--,--) \\
$0.75$ & $4.74$ & $5.495\times 10^{-5}$ & $2.389\times 10^{7}$ & $465.9$ & $10.76$ & $47$ & $20$ & $2\times 10^{-3}$ & $0.15$ & $(385,512)$ & $({\rm KTE},0.994,-)$ \\
-- & -- & $5.482\times 10^{-5}$ & $2.390\times 10^{7}$ & $516.0$ & $11.41$ &$45$ & -- & -- & -- & $(97,512)$ & ({\rm GL},--,--) \\
$0.80$ & $4.60$ & $8.388\times 10^{-5}$ & $9.834\times 10^{6}$ & $302.4$ & $7.66$ & $53$ & $14$ & $2\times 10^{-3}$ & $0.13$ & $(257,426)$ & $({\rm JVH},5,0.24)$ \\
-- & -- & $8.394\times 10^{-5}$ & $9.824\times 10^{6}$ & $324.6$ & $8.40$ &$55$ & -- & -- & -- & $(97,512)$ & ({\rm GL},--,--) \\
\bottomrule
\end{tabular}
}
\label{tab:results}
\end{table*}

\section{Temperature diffusion in a quasi-spherical shell with bottom 
topographic changes}
\label{sec:app_topo_theo}

We solve for the temperature diffusion $\nabla^2 T(r,\vartheta,\varphi) = 0$ 
in a solid with a spherical upper boundary located at $r=r_o$ and a bottom 
boundary which features topographic changes attributed to heat flux
variations coming from the underlying fluid layer. In a time-averaged sense, 
the latter boundary is described by the mean melt radius 
$\overline{\rmelt}(\vartheta,\varphi)$.
We assume fixed temperature at both boundaries with
$T(r=r_o,\vartheta,\varphi)=0$ and 
$T[r=\rmelt(\vartheta,\varphi),\vartheta,\varphi]=\tmelt$.
Let's introduce the following spherical harmonics expansions for the 
temperature and melt radius
\begin{equation}
\begin{aligned}
\overline{T}(r,\vartheta,\varphi)& =T_0(r)+ \sum_{\ell=1}^{+\infty}
\sum_{m=-\ell}^\ell \overline{T_{\ell m}}(r) 
Y_{ \ell m}(\vartheta,\varphi), \\
 \overline{\rmelt}(\vartheta,\varphi)& =\rmeltmean + 
 \sum_{\ell=1}^{+\infty}
\sum_{m=-\ell}^\ell
\overline{\xi_{\ell m}} Y_{\ell m}(\vartheta,\varphi),
\end{aligned}
\end{equation}
where $Y_{\ell m}$ is the spherical harmonic of degree $\ell$ and order $m$.
Assuming that the topographic changes are small compared to the mean 
melt radius $\rmeltmean$ \citep[an approximation also considered by][]{Kvorka24}, 
the boundary condition at the solid-liquid interface, held 
at a constant temperature $\tmelt$, can be approximated by

\begin{equation}
\begin{aligned}
 \tmelt & = \overline{T}(\rmelt,\vartheta,\varphi) \\
 & \approx  
\overline{T}(\rmeltmean,\vartheta,\varphi)+ \sum_{\ell\neq 0, m}
\overline{\xi_{\ell m}} Y_{ \ell m}(\vartheta,\varphi) \dfrac{\mathrm{d} 
T_0}{\mathrm{d}
r}(r=\rmeltmean)\,. \\
&\approx T_0(\rmeltmean) + 
\sum_{\ell\neq0,m}\left[T_{ \ell m}(\rmeltmean)+\overline{\xi_{\ell m}}
\dfrac{\mathrm{d} T_0}{\mathrm{d} r}(\rmeltmean)\right]
Y_{\ell m}(\vartheta,\varphi)\,,
\end{aligned}
\end{equation}
where the quadratic terms that would involve products of spherical harmonics 
expansions have been neglected \citep[for a similar approach applied to magnetic diffusion,
see][]{Styczinski22}.

At the leading order, $T_0$ is the solution of 

\begin{equation}
 \dfrac{1}{r^2}\dfrac{\mathrm{d} }{\mathrm{d} r}\left(r^2\dfrac{\mathrm{d} 
T_0}{\mathrm{d} r}\right)=0, \quad T_0(r=r_o)=0,\ T_0(r=\rmeltmean)=\tmelt,
\end{equation}
which yields

\begin{equation}
 \dfrac{\mathrm{d} T_0}{\mathrm{d} r}(r=\rmeltmean)=-\dfrac{\tmelt r_o 
}{\hS \rmeltmean},\ \hS=r_o-\rmeltmean,
\end{equation}
The first order terms are solution of

\begin{equation}
\begin{aligned}
 \dfrac{1}{r^2}\dfrac{\mathrm{d} }{\mathrm{d} r}\left(r^2\dfrac{\mathrm{d} 
\overline{T_{\ell m}}}{\mathrm{d} 
r}\right)-\dfrac{\ell(\ell+1)}{r^2}\overline{T_{\ell m}} &= 0, 
\quad  \forall (\ell, m) \\ 
T_{\ell m}(r=r_o)& =0, \\ T_{\ell m}(r=\rmeltmean)& =-\overline{\xi_{\ell m}}
\dfrac{\mathrm{d} 
T_o}{\mathrm{d} r}(r=\rmeltmean)\,.
\end{aligned}
\end{equation}
We seek for solution of the form $\overline{T_{\ell m}}(r) = \alpha r^\ell 
+\beta 
r^{-\ell-1}$. From the above boundary conditions, one gets
\begin{equation}
\beta = -\alpha r_o^{2\ell+1},\ 
 \alpha = \dfrac{\tmelt r_o \overline{\xi_{\ell m}} \rmeltmean^{\ell} 
}{\hS\left(\rmeltmean^{2\ell+1}-r_o^{2\ell+1}\right)  }\,.
\end{equation}
This yields the following spherical harmonic coefficients of the temperature 
field in the solid phase
\begin{equation}
 \overline{T_{\ell m}}(r)=\dfrac{\tmelt r_o}{\hS}
 \dfrac{\overline{\xi_{\ell m}} \rmeltmean^{\ell} 
}{\rmeltmean^{2\ell+1}-r_o^{2\ell+1}}\left(
r^\ell-\dfrac{r_o^{2\ell+1}}{r^{\ell+1}}\right),\ \forall(\ell,m)\,.
\end{equation}
From this expression, one can compute the temperature gradient along the 
solidus, again neglecting the quadratic terms. Introducing

\begin{equation}
 \partial \mathcal{Q}(\rmelt,\vartheta,\varphi) = \dfrac{\dfrac{\partial 
\overline{T}}{\partial 
r}(\rmelt,\vartheta,\varphi)-\dfrac{\mathrm{d} T_0}{\mathrm{d r}}({ 
\rmeltmean})}{%
\dfrac{\mathrm{d} T_0}{\mathrm{d r}}(\rmeltmean)},
\end{equation}
one gets

\begin{equation}
\partial \mathcal{Q} \approx \sum_{\ell\neq 0, m} 
\dfrac{\overline{\xi_{\ell m}}}{\rmeltmean}
\dfrac{\ell-1+(\ell+2)\eta_S^{2\ell+1}}{1-\eta_S^{2\ell+1}}
Y_{\ell m}(\vartheta,\varphi)\,,
\end{equation}
where $\eta_S \equiv \rmeltmean/r_o$ is the radius ratio of the solidus.

\bibliographystyle{cas-model2-names}

\end{document}